\documentclass[aps,pra,superscriptaddress]{revtex4}
\usepackage[intlimits]{amsmath}
\usepackage{amsfonts}
\usepackage{psfrag}
\usepackage{subfigure}
\usepackage[usenames]{color}
\usepackage{ifpdf}

\ifpdf
\usepackage{epstopdf}
\usepackage[pdftex]{hyperref}
\else
\usepackage[hypertex,ps2pdf]{hyperref}
\fi

\advance\voffset by -1cm
\advance\hoffset by -0.5cm
\newcommand\EQ {\begin{equation}}
\newcommand\EN {\end{equation}}
\newcommand\be            {\begin{equation}}
\newcommand\bea           {\begin{equation}\begin{array}l\displaystyle}
\newcommand\ee            {\end{equation}}
\newcommand\bes           {\begin{subequations}}
\newcommand\esu           {\end{subequations}}
\newcommand\erf[1]        {\eqref{#1}}
\newcommand\labl[1]       {\label{#1}\ee}
\newcommand{\st}{\stackrel}
\newcommand{\ud}{\mathrm d}
\newcommand{\eea}{\end{eqnarray}}

\newcommand{\bigx}[1]{\bBigg@{#1}}

\newcommand\eps           {\varepsilon}
\newcommand\fii           {\varphi}

\newcommand\mc            {\mathcal}

\newcommand\p            {\partial}
\newcommand\psid         {\psi^{\dagger}}
\renewcommand\th         {\theta}

\newcommand{\ket}[1]{\left\vert#1\right\rangle}
\newcommand{\bra}[1]{\left\langle#1\right\vert}

\renewcommand\vec[1]{{\vert{#1}\rangle}}

\newcommand\vev[1]{{\langle#1\rangle}}

\newcommand\no[1]{{\,:\!#1\!:\,}}

\def\3pt#1#2#3{{\langle{#1}\vert{#2}\vert{#3}\rangle}}

\newcommand\arxiv[2]      {\href{http://arXiv.org/abs/#1}{#2}}
\newcommand\doi[2]        {\href{http://dx.doi.org/#1}{#2}}

\begin{document}

\title{Integrability, Non-Integrability and Confinement}

\author{G. Mussardo}
\affiliation{SISSA and INFN, Sezione di Trieste, via Beirut 2/4, I-34151, 
Trieste, Italy}
\affiliation{International Centre for Theoretical Physics (ICTP), 
I-34151, Trieste, Italy}

\begin{abstract}
\noindent
We discuss the main features of quantum integrable models taking the classes of universality of the Ising model and the repulsive Lieb-Liniger model as paradigmatic examples. We address the breaking of integrability by means of two approaches, the Form Factor Perturbation Theory and semiclassical methods. Each of them has its own advantage. Using the first approach, one can relate the confinement phenomena of topological excitations to the semi-locality of the operator which breaks integrability. 
Using the second approach, one can control the bound states which arise in each phase of the theory and predict that their number cannot be more than two.  
\end{abstract}
\maketitle

\section{Introduction}
\label{sec:intro}
\noindent
Important developments of experimental tools for studying the behavior of quantum matter in low dimensions have triggered a renewed interest in integrable and non-integrable systems. Here I would like to mention, in particular, two recent experiments: 
\begin{itemize} 
\item the experiment on one-dimensional quantum Ising model, realized by means of CoNb$_2$O$_6$ (cobalt niobate) \cite{Coldea} and used to probe the $E_8$ symmetry of both the excitations \cite{Zam} and the correlation functions \cite{MusDel,DelSim} of the system. Such a system was also used to study the phenomena of confinement of topological excitations \cite{DMS,DM,McCoy} which occur once the degeneracy of the vacua is broken by the non-integrable deformation of the magnetic field. The quantum Ising model (and its tricritical version) will be in the following our paradigmatic models for discussing some of the theoretical predictions coming from quantum integrability and its breaking.  
\item the experiments on one-dimensional Lieb-Liniger gas, realized by using an optical trap in cold boson gas \cite{paredes04,kinoshita04,kinoshita06,vandruten08,nagerl}. In the paper \cite{kinoshita06} the authors reported, in particular, the preparation of out-of-equilibrium arrays of trapped one-dimensional (1D) Bose gases, each containing from 40 to 250 Rb$^{87}$ atoms, which did not noticeably equilibrate even after thousands of collisions. The result of this experiment has recently triggered a lot of interest on the role played by integrability in the off-equilibrium dynamics of a quantum system (for a review on the subject, see \cite{Silva} and references therein). Notice that, although the Bethe Ansatz solution of the Lieb-Liniger model is now almost fifty year old \cite{LL,yang}, the computation of its correlation functions at equilibrium has remained for long time an open problem and only recently a general approach has been set up for computing such quantities at zero and at a finite temperature \cite{KMT}. The Lieb-Liniger model will be our paradigmatic example to illustrate the efficiency of quantum field theory methods in the computation of correlation functions even in non-relativistic integrable models.
\end{itemize} 
The best way to characterize a quantum integrable system is through the non-diffractive scattering of its quasi-particle excitations \cite{sutherland}. This approach, closely related to the continuum limit of lattice models, directly leads to the formalism of integrable quantum field theory \cite{GMbook}.  It is worth to point out that the adoption of the continuum formalism of field theory (integrable or not-integrable) is not only extremely advantageous from a mathematical point of view but, in the scaling region nearby the critical points, is also perfectly justified from a physical point of view: in such a region the correlation length is larger than any other microscopic scale  and therefore a universal behavior is expected to emerge. It is such a universal behavior that is caught by the field theory. Moreover, a quantum field theory embodies a strong set of constraints coming from the compatibility of quantum mechanics with special relativity: this turns into general relations, such as the completeness of the multiparticle states or the unitarity of their scattering processes, which are extremely useful to successfully pin down the computation, say, of exact matrix elements of the order parameters and the universal shape of the correlation functions.  
 
The layout of the paper is as follows. In Section 2 we briefly recall the main properties of quantum integrable systems, using the class of universality of the Ising model and the Lieb-Liniger models as basic examples of the formalism. Non-integrable models will be the subject of the remaining sections and we will present two approaches which have been developed in recent years to study the phenomena accompanying the breaking of integrability, such as the confinement of the topological excitations, the decay of particles or the nature of the energy spectrum around different vacua. In Section 3 we will discuss some features of non-integrable models. In Section 4 we present the basic formulas of Form Factor Perturbation Theory, based on the exact matrix elements of the integrable models; in Section 5 we will present the main results coming from the Semiclassical Method, in particular those relative to the spectrum of neutral excitations present in each phase of the system. Our conclusions can be found in Section 6. 

\section{Quantum Integrable Models} 
Our prototype models for the discussion on the main features of quantum integrable systems are the Ising model and the Lieb-Liniger model (for recent reviews of these models see \cite{delfinorev} and \cite{weiss} respectively and references therein). As it will become clear below, the key quantity for the solution of these models is their $S$-matrix for this quantity permits to identify the spectrum of their excitations, to compute their correlation functions and to determine their thermodynamics properties.   

\subsection{Ising model.} 
Consider the Hamiltonian of the one-dimensional quantum spin chain 
of the Ising model
\begin{equation}
H \,=\,\sum_{i} \sigma^z_i \sigma^z_{i+1} + \lambda \sum_{i} \sigma^x_i + 
h \sum_{i} \sigma^z_i \,\,\,, 
\label{quantumIsing}
\end{equation}
where $\sigma_i^a$ are the Pauli matrices. This model is well known \cite{Sachdev} to have a phase transition at $h=0$ and $\lambda_c = 1$, the latter quantity related, in the corresponding 2d classical Ising model, with the critical temperature $T_c$.  At criticality, the Ising model is described by the first unitary minimal models of Conformal Field Theory (CFT) \cite{BPZ}. CFT permits to organize the operator content of the continuum limit of the model and to set the correspondence  
\begin{equation} 
\sigma_i^x \rightarrow \epsilon(r) 
\,\,\,\,\,\,\,
,
\,\,\,\,\,\,\,
\sigma_i^z \rightarrow \sigma(r) 
\label{operatorcontentIsing}
\end{equation}
where $\epsilon(r)$ and $\sigma(r)$ are the primary fields of CFT, identified respectively with the energy and magnetization densities. Hence, in the scaling region nearby the critical point, the Hamiltonian (\ref{quantumIsing}) can be written as  
\begin{eqnarray}
H & \,=\, & \left(\sum_{i} \sigma^z_i \sigma^z_{i+1} + \sum_{i} \sigma^x_i \right)+ 
(\lambda-1) \sum_i \sigma_i^x + 
h \sum_{i} \sigma^z_i \, \label{continuumHamiltonian}\\
& \simeq & 
H_{CFT} + \tau \int dr \,\epsilon(r) + h \int dr \,\sigma(r) \nonumber
\end{eqnarray} 
where $\tau = (\lambda -1)$ is the displacement from the critical value. 

\subsubsection{Integrable deformations}
The final expression of the Hamiltonian (\ref{continuumHamiltonian} ) is particularly convenient for discussing the nature of the various theories obtained by varying the coupling constants. Using the null-vector structure of CFT, it is possible to show that one gets an integrable model only in two cases \cite{Zam}:
\begin{eqnarray*} 
&& \tau\neq 0 \,\,\,\mbox {and} \,\,\,h=0, \mbox{denoted hereafter as {\em thermal deformation}}; \\
&& \tau=0 \,\,\,\mbox{and} \,\,\,h\neq 0,  \mbox{denoted hereafter as {\em magnetic deformation}}.
\end{eqnarray*} 
When both couplings are different from zero, the null-vector structure of CFT excludes instead the existence of any conserved charges and therefore the system is non-integrable \cite{GMreport}. The integrability of the two cases above means that in both thermal and magnetic deformation there is an infinite number of conserved charge ${\mathcal Q}_s$ of spin $s$, although the charges and their spin are different in the two cases: in particular, for the (infinite) spectrum of the spins of the conserved charges we have \cite{Zam}
\begin{equation}
\begin{array}{ll}
s =1,\,\,\,(\mbox{mod} \, 2) \,\,\,\, &\mbox{for the thermal deformation};\\
s = 1,7,11,13,17,19,23,29, \,\,\,(\mbox{mod} \,30) \,\,\,\, &\mbox{for the magnetic deformation}.
\end{array}
\end{equation}
The first values of the spins can be identified with the Coxeter exponents of the algebra 
$SU(2)$ and $E_8$ respectively: their appearance is related to the possibility of having alternative coset constructions of the Ising model at the critical point (for this point see, for instance, \cite{GMbook}). 

The existence of an infinite set of conserved charges ensures the elasticity and the factorization of the scattering processes of the excitations. Moreover, these conserved charges strongly constraint the spectrum of the excitations, whose number and mass can be obtained by closing the bootstrap of the $S$-matrix \cite{Zam}. This approach consists of the following steps: 
\begin{itemize}
\item parameterizing the dispersion relations of the particle $a$ by $E_a=m_a \cosh \theta_a, P_a=m_a \sinh\theta_a$ (where $\theta$ is the rapidity), the two-body elastic scattering amplitude $S_{ab}(\theta_{ab})$ relative to the process $a \times b \rightarrow a\times b$ depends, for the relativistic invariance, on the difference of the rapidity $\theta_{ab} = \theta_a - \theta_b$ and satisfies the unitary and crossing relations\footnote{We assume that the spectrum of a single particle is not degenerate and that each particle coincides with its anti-particle.}
\begin{equation} 
S_{ab}(\theta) S_{ab}(-\theta) =1 
\,\,\,\,\,\,
, 
\,\,\,\,\,\,
S_{ab}(\theta) = S_{ab}(i \pi -\theta) \,\,\,.
\label{unitaryandcrossing}
\end{equation}
\item Poles in the amplitude $S_{ab}(\theta)$ signal the presence of bound states, either in the $s$ or in $t$ channels. If $\theta_{ab}=i u_{ab}^c$ is identified with the $s$-channel pole, the mass of the corresponding bound state is given by 
\begin{equation}
m^2_c = m^2_a + m^2_b + 2 m_a m_b \cos u_{ab}^c \,\,\,.
\label{mass}
\end{equation}
\item The bootstrap principle asserts that bound states are on the same footing of the asymptotic states. This principle has a series of important consequences. Firstly, if there is the pole of the particle $c$ in the scattering amplitude $S_{ab}(\theta)$, then there will be the pole relative to the particle $a$ in the scattering amplitude $S_{bc}(\theta)$, as well as there will be the pole relative to the particle $b$ in $S_{ac}(\theta)$. Secondly, 
the position of the poles in the three channels satisfies 
\begin{equation}
u_{ab}^c + u_{bc}^a + u_{ac}^b = 2 \pi \,\,\,.
\end{equation}
and, finally, the amplitudes are related by the bootstrap equation ($\bar u_{lm}^n \equiv \pi - u_{lm}^n$)
\begin{equation}
S_{ab}(\theta) = S_{ac}(\theta + i \bar u_{cb}^d) S_{ad}(\theta - i \bar u_{bd}^c)  \,\,\,.
\label{bootstrap}
\end{equation}
\end{itemize}
So, to summarize the rules of the game, in order to determine the S-matrix (and the spectrum) of an integrable theory by means of the bootstrap approach one has to find a set of poles relative to all amplitudes $S_{ab}$ which can be interpreted in terms of bound states of the asymptotic particles and which are compatible with the bootstrap equation (\ref{bootstrap}). The masses of the particles are determined by the relation (\ref{mass}). In practice all this means starting from the amplitude that involves the lighest particle, therefore with the simplest pole structure, and then iteratively applying the bootstrap equations (\ref{bootstrap}) to get the scattering amplitudes involving the bound states of higher mass. Let's see how this formalism works for the two integrable deformations of the Ising model. 

{\bf Thermal deformation}. The thermal case is known to correspond to a theory of a free Majorana fermion of mass $m$ (proportional to $\tau$), with two-body $S$-matrix  simply given by $S=-1$. Such a theory does not have additional bound states. 

{\bf Magnetic deformation}. The magnetic case is more subtle. The set of conserved charges implies the existence of a lowest mass particle $A_1$ (which is bound state of itself), plus at least two additional particles $A_2$ and $A_3$; moreover, also thanks to the conserved charges, it is also possible to pin down the location of their poles in the amplitude $S_{11}(\theta)$. With all these data, the amplitude $S_{11}(\theta)$ is given by \cite{Zam}
\begin{equation}
S_{11}(\theta) = \st{\bf 1}{\left(\frac{2}{3}\right)} \, \st{\bf 2}{\left(\frac{2}{5}\right)} \, 
\st{\bf 3}{\left(\frac{1}{15}\right)} \,\,\,,
\label{fundamental}
\end{equation}
where 
\[
(x) \equiv \frac{\tanh\frac{1}{2}(\theta + i \pi x)}{
\tanh\frac{1}{2}(\theta-i\pi x)}\,\,\,.
\]
The pole at $\theta=i \frac{2\pi}{3}$ corresponds to the particle $A_1$ itself, $\theta = i \frac{2 \pi}{5}$ corresponds to the particle $A_2$ whereas the one at $\theta = i \frac{\pi}{15}$ to the particle $A_3$. If $m_1$ is the mass of the particle $A_1$, the mass of these new particles are $m_2 = 2 m_1 \cos\frac{\pi}{5}$ and 
$m_3 = 2 m_1 \cos\frac{\pi}{30}$, respectively. 

Applying the boostrap equation relative to the poles, one could get the amplitudes $S_{12}(\theta)$ and $S_{13}(\theta)$, where other poles appear in correspondence to new particles.  Applying the bootstrap equations also to these poles and carrying on all the necessary steps to close consistently the process (i.e. a consistent interpretation of all pole structure of the amplitudes in terms of the identified particles), A. Zamolodchikov showed that the bootstrap procedure closes with 36 amplitudes and 8 particles, whose exact mass spectrum is given by \cite{Zam}    
 \begin{eqnarray}
m_1 &=& m \nonumber \\
m_2 &=& 2 m_1 \cos\frac{\pi}{5} = (1.6180339887..) \,m_1\nonumber\\
m_3 &=& 2 m_1 \cos\frac{\pi}{30} = (1.9890437907..) \,m_1\nonumber\\
m_4 &=& 2 m_2 \cos\frac{7\pi}{30} = (2.4048671724..) \,m_1\nonumber \\
m_5 &=& 2 m_2 \cos\frac{2\pi}{15} = (2.9562952015..) \,m_1\label{E8spectrum}\\
m_6 &=& 2 m_2 \cos\frac{\pi}{30} = (3.2183404585..) \,m_1\nonumber\\
m_7 &=& 4 m_2 \cos\frac{\pi}{5}\cos\frac{7\pi}{30} = (3.8911568233..) \,m_1\
\nonumber\\
m_8 &=& 4 m_2 \cos\frac{\pi}{5}\cos\frac{2\pi}{15} = (4.7833861168..) \,m_1
\nonumber
\end{eqnarray}
Notice that the mass of the particles $A_4, \ldots, A_8$ are above the mass threshold 
$2 m$: in a generic theory these particles would be all unstable whereas here their stability is ensured by the integrability of the theory.

\subsubsection{ Form Factors and Correlation Functions}

\begin{figure}[t]
\centerline{\scalebox{0.5}
{\includegraphics{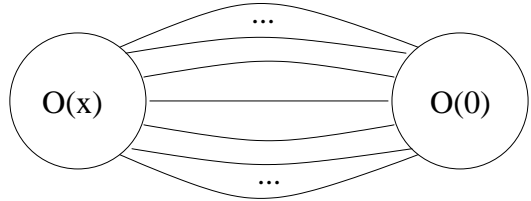}}}
\caption{{\em Spectral representation of the two-point correlation functions.}}
\label{organetto}
\end{figure}

In addition to the exact expression of the $S$-matrix, in quantum integrable theories  one can also compute the exact matrix elements of the order parameters on the  asymptotic state (the so-called {\em Form Factors}) \cite{karowski, smirnov}
\begin{equation}
F^{\Phi}_{a_1,a_2,\ldots,a_n} (\theta_{a_1},\theta_{a_2},\ldots,\theta_{a_n}) \,=\,
\langle 0 |\Phi(0) | A_{a_1}(\theta_{a_1}) A_{a_2}(\theta_{a_2}) \ldots A_{a_n}(\theta_{a_n}) \rangle \,\,\,.
\label{FORMFACTORS}
\end{equation}
The Form Factors satisfy a set of functional and recursive equations based on unitarity, crossing, factorization and pole structure of the $S$-matrix: different solutions of these equations identify the operator content of the theory \cite{CardyMussardo}. We refer the reader either to the original references \cite{karowski,smirnov} for the key equations of the Form Factor approach or to the book \cite{GMbook} for a general introduction to this formalism. Below we will simply quote the main results relative to the two integrable deformations of the Ising model.  Notice that, once we have determined the form factors of a given operator, its correlation functions can be written in terms of the spectral representation series using the completeness relation of the multi-particle states. For instance, the two-point correlation function of the operator ${\mathcal O}(x)$ (in the euclidean space) can be written as \footnote{To simplify the expression, we assume that there is only one kind of excitation}
\begin{eqnarray}
& & \langle{\mathcal O}(x)\,{\mathcal O}(0)\rangle\,=
\sum_{n=0}^{\infty}
\int \frac{d\theta_1\ldots d\theta_n}{n! (2\pi)^n}
<0|{\mathcal O}(x)|\theta_1,\ldots,\theta_n>_{\rm in}{}_{\rm in}
<\theta_1,\ldots,\theta_n|{\mathcal O}(0)|0>
\label{correlation} \nonumber \\
& &\hspace{3mm} =\,\sum_{n=0}^{\infty}
\int \frac{d\theta_1\ldots d\theta_n}{n! (2\pi)^n}
\mid F_n(\theta_1\ldots \theta_n)\mid^2 \exp \left(-mr\sum_{i=1}^n\cosh\theta_i
\right)
\end{eqnarray}
where $r$ is the radial distance $r=\sqrt{x_0^2 + x_1^2}$ (Figura \ref{organetto}). Similar expressions, although more complicated, hold for the $n$-point correlation functions. 
It is worth stressing some advantages of this method.  
\begin{itemize}
\item The integrals present in the spectral series are all convergent, in sharp contrast with the formalism based on the Feynman diagrams, in which one encounters the divergences of the various perturbative terms. The reason is that the form factors employ {\em ab initio} all the {\em physical} parameters of the theory and therefore the divergences of the perturbative series (which, on the contrary, refers to the {\em bare} parameters) are absent.   
\item 
If the correlation functions do not have ultraviolet singularities particularly violent (this is the case of the correlation functions of the relevant fields), the spectral series has an extremely fast convergent behavior for all values of  $m r$. In the infrared region, i.e. for large values of $m r$, this is pretty evident from the nature of the series, because its natural parameter of expansion is $e^{-m r}$. The fast convergent behavior also in the ultraviolet region $m r \rightarrow 0$ has instead a twofold reson: the peculiar behavior of the $n$-particle phase space in two-dimensional theories, further enhanced by the form factors behavior, as clarified in \cite{GMbook,CARDYMUSSARDOO3}. 
\end{itemize}

\noindent
{\bf Form Factors of the thermal deformation}. 
The field theory associated to the thermal deformation of the Ising model is known to be self-dual for the exchange $\tau \rightarrow - \tau$: the energy operator changes sign under this transformation, $\epsilon(x) \rightarrow - \epsilon(x)$, whereas the order parameter $\sigma(x)$ goes into the disorder operator $\mu(x)$, i.e. $\sigma \rightarrow \mu$. For $\tau > 0$,  $\sigma(x)$ is a $Z_2$ odd operator with non-zero fermion number while $\mu(x)$ is a $Z_2$ even operator with zero fermion number: hence, 
$\sigma(x)$ has non-zero matrix elements only on odd number of fermions,
while $\mu(x)$ has non-zero matrix elements only on even number of fermions.  
For $\tau < 0$, the role of the two operators is swapped and the situation 
is reversed: this symmetry is due to the self-duality of the model. 
For $\tau > 0$, the matrix elements of the various operators (between the vacuum state $\ket{0}$ and the asymptotic states of $n$ fermions identified 
by their rapidities $\ket{\theta_1,\theta_2 ,\ldots,\theta_n}$) 
are given by \cite{Alyosha}   
\[
\bra{0} \epsilon(0,0) \ket{\theta_1,\ldots,\theta_n} = 
\left\{  \begin{array}{cl}
i m \, \sinh\frac{\theta_1 - \theta_2}{2} & {\rm for} \; n = 2 \\ 0 & {\rm otherwise} 
\end{array} \right.
\]
\[
\bra{0} \sigma(0,0) \ket{\theta_1,\ldots ,\theta_{2n+1}} = (i)^{2n+1}
\prod_{i < j}^{2n +1} \tanh\frac{\theta_i-\theta_j}{2} 
\]
\[ 
\bra{0} \mu(0,0) \ket{\theta_1,\ldots ,\theta_{2n}} = (i)^{2n}
\prod_{i < j}^{2n } \tanh\frac{\theta_i-\theta_j}{2} \;.
\]
Inserted into the spectral representation, the energy form factors give the (euclidean) correlation function $\langle \epsilon(x) \epsilon(0)\rangle = m^2 (K_1^2(m x) - K_0^2(mx))$ (where $K_i(y)$ are the modified Bessel functions). The magnetization form factors give rise instead to the Painleve' equation satisfied by the correlators 
$\langle \sigma(x) \sigma(0)\rangle$ and $\langle \mu(x) \mu(0) \rangle$ \cite{mccoypainleve}. Notice that the matrix elements of the magnetization operators have a pole at $\beta_i - \beta_j = i \pi$. This observation will be important when we discuss the Form Factor Perturbation Theory relative to the non-integrable deformation of the thermal theory. 

\noindent
{\bf Form Factors of the magnetization deformation}. As mentioned above, 
the Ising model in a magnetic field has quite a rich $S$-matrix, with amplitudes that have poles up to 12th order. In addition to the functional and recursive equations, the form factors of this theory also satisfy other recursive equations related to the higher poles of the $S$-matrix: the relative formulas and the final exact expressions of the Form Factors can be found in the papers \cite{MusDel,DelSim}. Here we only report the 
exact result relative to the one-particle amplitudes $Z_n$ of the spin-spin correlation function
\EQ
\langle \sigma(q) \sigma(-q) \rangle \,=\,
\frac{Z_1}{q^2+m_1^2} + \frac{Z_2}{q^2+m_2^2} + \cdots + 
\frac{Z_8}{q^2+m_8^2} + \cdots 
\EN 
They are given by the square of the one-particle Form Factors  
\EQ
Z_k\,=\,|\langle 0 |\sigma(0) |A_k(\theta)\rangle|^2\,\,\,.
\EN 
Their actual value depends on the normalization of the operator $\sigma(x)$ but universal numbers can be extracted in terms of the ratios $\tilde Z_k = Z_k/Z_1$. Their values is in Table 1 and their plot is given in Figure 2. Figure 2 has to be compared with the experimental data reported in Figure 4E of the paper \cite{Coldea} and one could easily see the two plots agree one to the other.   
\begin{figure}[t]
\centerline{\scalebox{0.5}
{\includegraphics{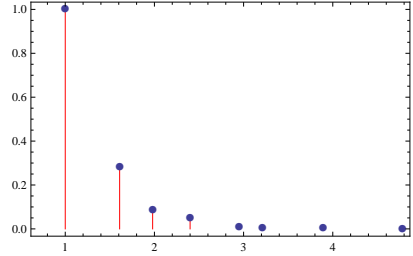}}}
\caption{{\em Quasi-particle weights of the spin-spin correlation function normalized to the weight of the particle with the lowest mass.}}
\label{weightsZ}
\end{figure}

\begin{table}[h]
\begin{center}
\begin{tabular}{|l|c|}\hline
$\tilde Z_1 $ & 1  \\
$\tilde Z_2 $ &  0.279242\\
$\tilde Z_3 $ &  0.0847955\\
$\tilde Z_4$ &  0.0496251\\
$\tilde Z_5 $ &  0.00885986\\
$\tilde Z_6$ &  0.00458335\\
$\tilde Z_7 $ &  0.000656846\\
$\tilde Z_8 $ & 0.0000224409 \\ \hline
\end{tabular}
\end{center}
\caption{Universal ratios made of one-particle amplitudes in spin-spin correlator.}
\end{table}

\subsection{Lieb-Liniger model}

Our second example of integrable model is the Lieb-Liniger (LL) model \cite{LL} that 
describes the low-temperature properties of one-dimensional interacting non-relativistic Bose gases. Its accurate experimental realization \cite{paredes04,kinoshita04,kinoshita06,vandruten08,nagerl} has stimulated a lot of interest and has opened new perspectives in the field of strongly correlated systems. Thanks to the highly controllable experimental set-ups, it is possible nowadays to thoroughly investigate questions of general nature concerning quantum extended
systems, such as the dynamics of integrable systems in the presence of
small non-integrable perturbations (e.g., three-body interactions and/or a weak external trapping potential), the issue of thermalization in quantum integrable and non-integrable systems and the behavior of various susceptibilities and response functions.

Following the recent papers \cite{KMT}, we will briefly discuss how one can compute the correlation functions of this model (at finite density and at finite temperature) using its relation to the relativistic integrable Sh-Gordon model. The LL Hamiltonian for $N$ interacting bosons of mass $m$ in one dimension is 
\be
\label{HAM_LL}
H\,=\,-\frac{\hbar^2}{2m}\sum_{i=1}^N\frac{\p^2}{\p x_i^2}+2\lambda\,
\sum_{i<j}\delta(x_i-x_j)\,, 
\ee
and the corresponding (non-relativistic) field theory is given by the quantum non-linear Schr\"odinger model \cite{korepin}, that employs the complex field $\Psi$ and the Lagrangian  
\be
{\cal L}= -\frac{\hbar^2}{2m}|\nabla\psi|^2
+ 
i\,\frac\hbar2\left(\psid\frac{\p\psi}{\p t} - \frac{\p\psid}{\p
  t}\psi\right) - \lambda\,|\psi|^4\,.
\label{LagrangianNLS}
\ee
The effective coupling constant of the LL model is given by the dimensionless parameter $\gamma=2m\lambda/\hbar^2n$, where $\lambda > 0$ is the coupling entering the Hamiltonian (\ref{HAM_LL}) while $n=N/L$ is the density of the gas ($L$ is the length of the system). It is convenient to express the values of temperature in units of the temperature $T_D=\hbar^2 n^2/2mk_B$ of the quantum degeneracy, $\tau=T/T_D$. The non-relativistic two-body elastic S-matrix of the LL model is \cite{LL,yang}
\be
S_\text{LL}(p,\lambda)\,=\,\frac{p-i 2m \lambda/\hbar}{p+i 2m \lambda/\hbar}\,,
\label{eq:SLL}
\ee
where $p$ is the momentum difference of the two particles. Consider now the Sh-Gordon model in (1+1) dimensions, i.e.\ the integrable and relativistic  invariant field theory defined by the Lagrangian 
\be
\mc{L}_\text{ShG}= \frac12\left[\left(\frac{\p\phi}{c\,\p
  t}\right)^2-\left(\nabla\phi\right)^2\right] -
\frac{\mu^2}{g^2}
\cosh(g\phi)\,,
\label{LagrangianShG}
\ee
where $\phi=\phi(x,t)$ is a real scalar field, $\mu$ is a mass scale and $c$ is the speed of light. The parameter $\mu$ is related to the physical (renormalized) mass $m$ by $\mu^2=\pi \alpha m^2c^2 / \hbar^2 \sin(\pi\alpha)$, where $\alpha=\hbar c\,g^2 / (8\pi+\hbar c\,g^2)$ \cite{FMS}. Let's express the dispersion relation of the energy $E$ and the momentum $P$ of a particle as $E=m c^2 \cosh\th$, $P=m c \sinh\th$, where $\th$ is the rapidity and $c$ the light speed. The Sh-Gordon is an integrable model and therefore all its scattering processes are purely elastic and can be factorized in terms of the two-body S-matrices \cite{Arisha}
\be
S_\text{ShG}(\th,\alpha)\,=\,\frac{\sinh\th-i\,\sin(\alpha\pi)}{\sinh\th+
i\,\sin(\alpha\pi)}\,,
\label{eq:SH-G}
\ee
where $\th$ is the rapidity difference of the two particles. The key observation is that taking simultaneously the non-relativistic and weak-coupling limits of the ShG model such that  
\be
g\to0,\;c\to\infty,\quad g\,c=4\sqrt{\lambda}/\hbar=\text{fixed}\,, 
\label{eq:limit}
\ee
its $S$-matrix (\ref{eq:SH-G}) becomes identical to the $S$-matrix (\ref{eq:SLL}) of the LL model. In this double limit procedure the coupling $\lambda$ does not need to be small, i.e.\ with this mapping we can study the LL model at arbitrarily large values of the dimensionless coupling $\gamma$. The mapping between the $S$-matrices of the two models suggests that this mapping should extend also to their Lagrangians and Thermodynamic Bethe Ansatz (TBA) equations. This is indeed true. In the non-relativistic limit, the real scalar field can be expressed as 
\[
\phi(x,t)=\sqrt{\frac{\hbar^2}{2m}}\left(\psi(x,t)\,
  e^{-i\frac{mc^2}\hbar\,t}+\psid(x,t) e^{+i\frac{mc^2}\hbar\,t}\right)\,,
\]
and, once inserted into the Lagrangian (\ref{LagrangianShG}), in the limit $c\to\infty$  one omits all oscillating terms. Moreover, when $g \to 0$ of eqn \erf{eq:limit} is considered, the $\psid\psi$ terms coming from the potential and kinetic parts of (\ref{LagrangianShG}) cancel each other, while all higher terms of the series expansion of the potential, but the quartic one, vanish. Hence, the ShG Lagrangian (\ref{LagrangianShG}) reduces to the non-linear Schr\"odinger Lagrangian (\ref{LagrangianNLS}). The commutation relation
$
[\phi(x,t),\Pi(x',t)] = i\hbar\,\delta(x-x')
$
implies for the non-relativistic operators
$
[\psi(x,t),\psid(x',t)]=\delta(x-x')\,.
$

Similarly the TBA equations of the ShG model (given for instance in \cite{klass}) reduce to the ones of the LL model, written down in \cite{yang}. In the LL model at a finite $T$,  the pseudo-energy $\epsilon(T,\mu)$ satisfies the non-linear integral equation 
\be
\eps(T,\mu)\,=\, \frac{p^2/2m -\mu}{k_B T} - \fii \circ 
\log\left(1+e^{-\eps}\right)\,,
\label{eq:TBA}
\ee
where $\mu$ is the chemical potential conjugated to the finite density $n$ of the gas, $\fii(p) = - i\frac\p{\p p}\log S_{LL}(p)$ is the derivative of the phase shift and $\fii \circ f \equiv \int_{-\infty}^\infty\frac{\mathrm{d} p'}{2\pi}\,\fii_(p-p') f(p')$. The solution of this integral equation leads to the free energy and to all other thermodynamical data of the model.   

In the light of the mapping between the $S$-matrix, the TBA and the operators of the LL and the Sh-Gordon models, we can now proceed to compute the expectation values of the LL model. At equilibrium the expectation value of an operator $\mc O = \mc O(x)$ at temperature $T$ and at finite density is given by 
\be
\vev{\mc O} = \frac{\mathrm{Tr}\left(e^{-(H-\mu N)/(k_\text{B}T)}\mc{O}\right)}
{\mathrm{Tr}\left(e^{-(H-\mu N)/(k_\text{B}T)}\right)}\,. 
\labl{eq:vev}
In a relativistic integrable model the above quantity can be neatly expressed as \cite{leclair} 
\be
\vev{\mc O} =\sum_{n=0}^\infty\frac1{n!}
\int_{-\infty}^\infty 
\left(\prod_{i=1}^n\frac{\ud\th_i}{2\pi} f(\th_i)
\right) 
\3pt{\overleftarrow{\th}}{\mc O(0)}{\overrightarrow{\th}}_\text{conn}\,,
\labl{eq:muss}
where $f(\th_i) = 1/(1+e^{\eps(\th_i)})$ and $\overrightarrow{\th} \equiv 
\th_1,\dots,\th_n$ ($\overleftarrow{\th} \equiv \th_n,\dots,\th_1$)
denote the asymptotic states entering the traces in (\ref{eq:vev}). This formula employs both the pseudo-energy $\epsilon(\th)$ and the connected diagonal form factor of the operator ${\mc O}$, defined as 
$
\3pt{\overleftarrow{\th}}{\mc O}{\overrightarrow{\th}}_\text{conn}=
FP\left(\lim_{\eta_i\to0} \3pt{0}{\mc
  O}{\overrightarrow{\th},\overleftarrow{\th} - i\pi+i \overleftarrow{\eta}}\right) 
$
where $\overleftarrow{\eta} \equiv \eta_n,\dots,\eta_1$ and $FP$ in front of the expression means taking its finite part, i.e. omitting all the terms of the form $\eta_i/\eta_j$ and $1/\eta_i^p$ where $p$ is a positive integer. For a recent proof of this formula see \cite{Pozsgay}. 

To compute the expectation values of the LL model we have then to apply eqn (\ref{eq:muss}) according to the following steps: (a) solve the integral equation (\ref{eq:TBA}) for $\epsilon(\th)$; (b) identify the relevant form-factors of the ShG model; (c) take the non-relativistic limit of both the form-factors and eqn (\ref{eq:muss}). To proceed, let's assume we have solved numerically eqn (\ref{eq:TBA}) (a task that can be easily done) and therefore let's attack the last two points. The generic $m$-particle form factor of a local operator ${\mc O}$ in the ShG model can be written as \cite{FMS,mussardo}
\be
F_m^{\mc O}(\th_1,\dots,\th_m) = \,Q_m^{\mc O} (x_1,\dots,x_m)\,\prod_{i<j}\frac{F_\text{min}(\th_{ij})}{x_i+x_j}\,,
\ee
where $x_i=e^{\th_i}$ and $Q_m^{\mc O}$ are the  symmetric polynomials in the $x$'s that fully characterize the operator ${\mc O}$. The explicit expression of $F_\text{min}(\th)$ is given in \cite{FMS} but the only thing needed here is its functional equation 
\[
F_{\rm min}(i\pi+\th) F_{\rm min}(\th)=
\frac{\sinh\th}{\sinh\th+ i \sin(\pi \alpha)}\,.
\]
We are interested, in particular, in the symmetric polynomials $Q_m^{(q)}$ of the
exponentials ${\mc O}_q = e^{q g \phi}$ since, using their Taylor expansion, we can extract the form factors of all normal ordered operators $\no{\phi^k}$. Their close expression is \cite{mussardo} 
\be
Q_m^{(q)}= [q] \left(\frac{4\sin(\pi\alpha)}{N}\right)^{\frac{m}2}\det
M_m(q)\,,
\labl{eq:FFexp}
where $M_m(q)$ is an $(m-1)\times(m-1)$ matrix with elements $ \left[M_m(q)\right]_{i,j}=\sigma^{(m)}_{2i-j}[i-j+q]
$. Above, $[x]\equiv
\sin(x\pi\alpha)/\sin(\pi\alpha)$ while $\sigma^{(m)}_a$ ($a=0,1,\dots, m$) are the elementary symmetric polynomials in $m$ variables. 

\begin{figure}[t]
\centerline{
\scalebox{0.27}{\includegraphics{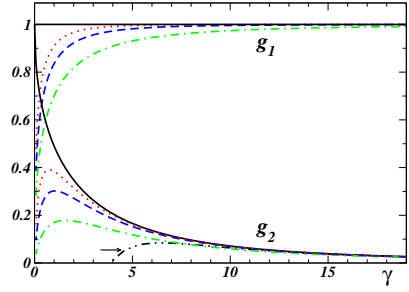}}
}
\caption{$g_1$ and $g_2$ using form factors up to 
$n=\textcolor{green}{4}$, $\textcolor{blue}{6}$ and $\textcolor{red}{8}$ particles, 
respectively with green dot-dashed,  
blue dashed and red dotted lines. The exact values are given by the 
solid lines whereas 
the dot-dashed line below, indicated by the arrow,  corresponds to the strong coupling 
expansion (\ref{strong-coupling}).}
\label{fig:fig1}
\end{figure}

We can now compute the local $k$-particle correlation function $g_k$ of the LL model defined by 
\be
\vev{\psid\,^k\psi^k}=n^k\,g_k(\gamma,\tau)\,,
\ee
where $k$ is an integer ($k=1,2,3,\dots$). The $g_k$'s are functions of the dimensionless LL coupling $\gamma$ and of the reduced temperature $\tau$. 
The relation between $g_k$ in the LL model and the corresponding quantity in the ShG model  in the limit (\ref{eq:limit}) is given by  
\[\vev{\no{\phi^{2k}}}\to 
\left(\frac{\hbar^2}{2m}\right)^k\binom{2k}{k}\vev{\psid\,^k\psi^k}\,.
\]
Using eqn \erf{eq:muss} and the connected form factors of the corresponding operator the final expression is 
\begin{gather}
\vev{\psid\,^k\psi^k} = 
 \binom{2k}{k}^{-1}\!\!\left(\frac{\hbar^2}{2m}\right)^{-k} 
\sum_{n=1}^\infty {\cal F}_n\,,
\label{eq:formula}\\
{\cal F}_n = \frac{1}{n!} \int_{-\infty}^{\infty} 
\left(\prod_{i=1}^n \frac{d p_i}{2\pi} f(p_i)\right)
\tilde F^{\no{\,\phi^k\,}}_{2n,\text{conn}}(p_1,\ldots,p_n)\,,\nonumber
\end{gather}
where
\[
\tilde
F^{\no{\,\phi^k\,}}_{2n,\text{conn}}(\{p_i\})=\lim_{c\rightarrow \infty,
g\rightarrow 0}\,\left(\frac1{mc}\right)^nF^{\no{\,\phi^k\,}}_{2n,\text{conn}}(\{\th_i=\frac{p_i}
{mc}\})
\]
are the double limit (\ref{eq:limit}) of the connected form factors. It is worth pointing out that the matrix elements of the LL models computed by the Algebraic Bethe Ansatz 
\cite{korepin} can be put in direct correspondence with the Form Factors of the relativistic Sh-Gordon model \cite{KMP}.
As shown in \cite{KMT}, the series (\ref{eq:formula}) are nicely saturated by the first few terms for sufficiently large values of $\gamma$ ($\gamma = 0$ is a singular point of the model
\cite{LL}, therefore one cannot expect a priori any fast convergence nearby). 
A first check of eqn (\ref {eq:formula}) is given by the case $k=1$: using (\ref{eq:formula}) (with a chemical potential $\mu$ that ensures the finite density $n$) and summing up the series, one easily checks that $\vev{\psid\psi}=n$ and $g_1=1$ (a result which comes from translational invariance). As shown in Fig.~\ref{fig:fig1}, the exact value $g_1=1$ (solid line) is rapidly approached by just the first terms of (\ref{eq:formula}): the convergence of the series is always remarkably fast for all $\gamma \geq 1.5$, where the exact value is obtained within a $5\%$ accuracy just using its first four terms. 
\begin{figure}[t]
\scalebox{0.23}{\includegraphics{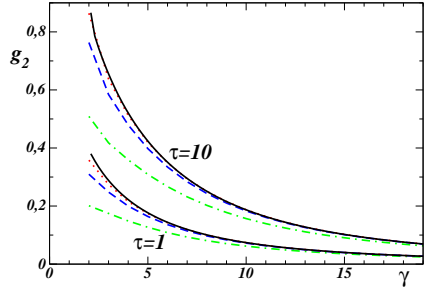}}
\caption{$g_2$ at $\tau=1$, $10$ using form factors up to $n=\textcolor{green}{4}$, 
$\textcolor{blue}{6}$ and $\textcolor{red}{8}$ particles with 
green dot-dashed, blue dashed and red dotted lines, respectively. The
solid lines show the exact result.}
\label{fig:fig2}
\end{figure}
Eq. \ref{eq:formula}) also permits to easily recover the leading order of the strong coupling (i.e.\ large $\gamma$) expansion of all $g_k$: this always comes from the first non-zero integral in the series \erf{eq:formula} and therefore  
\be
g_k=\frac{k!}{2^k}\left(\frac\pi\gamma\right)^{k(k-1)} I_k +\dots\,,
\label{eq:leading}
\ee
where 
$I_k=\int_{-1}^1\ud k_1\dots\int_{-1}^1\ud k_k \prod_{i<j}^k (k_i-k_j)^2$. This result coincides with the one obtained in \cite{gangardt}.

The quantity $g_2$ can be exactly determined via the Hellmann--Feynman theorem
\cite{kheruntsyan}: its plot at $T=0$ is shown in Fig.~\ref{fig:fig1} together with our determination from eqn \erf{eq:formula} and, as before, one observes also in this case a fast convergent behaviour of the series. The strong coupling regime
of $g_2$ can be computed by expanding in powers of $\gamma^{-1}$ all the terms in eqn (\ref{eq:formula}) and for $T=0$ we get 
\be
g_2=\frac43\frac{\pi^2}{\gamma^2}\left(1-\frac6\gamma+(24-\frac85\pi^2)\frac1{\gamma^2}\right)+\mc O(\gamma^{-5})\,,
\label{strong-coupling}
\ee
in agreement with the result of the Hellmann--Feynman theorem
\cite{gangardt,kheruntsyan}. Expression \erf{strong-coupling} is also 
plotted in Fig.~\ref{fig:fig1} in order to show that 
the determination of $g_2$ (at finite $\gamma$)  obtained from the first terms of eqn \erf{eq:formula} is closer to the exact result, because any of them contains
infinitely many powers of $\gamma$.  At finite temperatures the convergence of the series is also pretty good and the results are shown in Fig.~\ref{fig:fig2}. 
\begin{figure}[t]
\scalebox{0.39}{\includegraphics{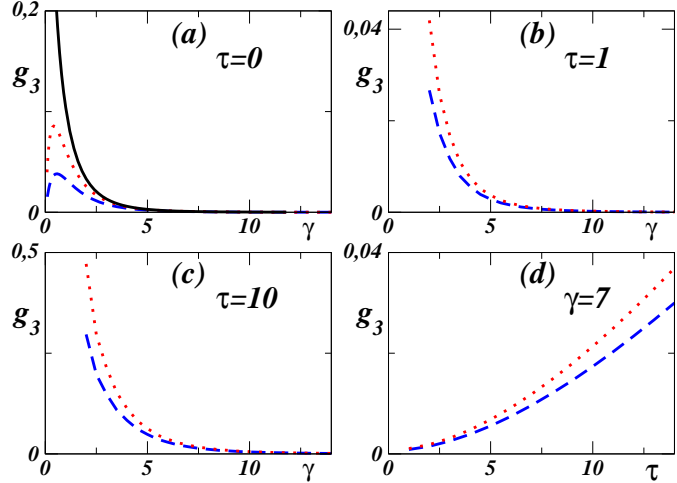}}
\caption{$g_3$ vs $\gamma$ at : (a) $\tau=0$,   (b) $\tau= 1$ and (c) $\tau=10$. In (d) 
we plot $g_3$ vs $\tau$ at $\gamma=7$. In all figures the blue dashed and the red 
dotted lines refer to $n=\textcolor{blue}{6}$ and 
$\textcolor{red}{8}$ particles, respectively. The solid line in (a) 
is the exact value of $g_3$ at $\tau=0$.}
\label{fig:fig3}
\end{figure}

A new important quantity get by the method above is $g_3$, a quantity known exactly at $T=0$ \cite{cheianov}, but only approximately at $T>0$ \cite{gangardt}. This quantity  
is related to the recombination rate of the quasi-condensate and thus to the lifetime of the experiments. Its strong coupling limit at $T=0$  is 
\be
g_3=\frac{16}{15}\frac{\pi^6}{\gamma^6}\left(1-\frac{10}\gamma \right)+\mc O(\gamma^{-8})\,. 
\ee
The plot of $g_3$ at $\tau=0$ using form factors up to $n=6$ and $8$ particles (i.e. one or two terms of the series \erf{eq:formula}) is in Fig.~\ref{fig:fig3}(a) and, as in previous examples, it shows a nice convergent pattern to the exact value found in \cite{cheianov}. Figs.\ \ref{fig:fig3}(b,c) show $g_3$ as a function of $\gamma$ at fixed temperature $\tau$, while Fig.~\ref{fig:fig3}(d) shows instead $g_3$ as a function of $\tau$ at a fixed value of $\gamma$.

In closing this section, it should be mentioned that the method discussed above works also in the case of the super Tonks-Girardeau gas \cite{superTonks}, i.e. a highly excited, strongly correlated state obtained in quasi one-dimensional Bose gases by tuning the scattering length to large negative values of $\gamma$ by using a confinement-induced resonance. 

\section{Non--Integrable Aspects}

Two--dimensional massive Integrable Quantum Field Theories (IQFTs) have proven to be one of the most successful topics of relativistic field theory: in addition to the examples discussed in the previous section, it has found a large variety of applications to many other models (see, for instance, \cite{GMbook}). However, despite the elegance and the undeniable success of these methods, the generic situation that occurs in many-body quantum physics is that of a non-integrable dynamics: many interesting models fall within this class and therefore it would be highly desirable to develop an appropriate formalism to deal with the lack of integrability.
The breaking of integrability is expected to considerably increase the difficulties of the mathematical analysis, since the scattering processes of the excitations are no longer elastic. Non--integrable field theories are in fact generally characterized by particle production amplitudes, resonance states and, correspondingly, decay events. All these features strongly effect the analytic structure of the scattering amplitudes, introducing a rich pattern of branch cut singularities, in addition to the pole structure associated to bound and resonance states. 

Among the many possible ways in which a model can be non-integrable, a particularly interesting situation occurs when we consider Hamiltonian defined in terms of a conformal Hamiltonian deformed by two relevant operators, each of them giving rise {\em individually} to an integrable model
\EQ
H \,=\,H_{CFT} + \lambda_1 \,\int d x \,\varphi_1(x) + 
\lambda_2 \,\int d x \,\varphi_2(x) \,\,\,. 
\label{multiple}
\EN 
There are in fact many interesting physical systems that belong to this class of non-integrable models. Let's briefly discuss two of them.  
\begin{itemize}
\item The first is the Ising model (\ref{continuumHamiltonian}) at $\lambda \neq 1$ in an external magnetic field $h$. We have previously seen that when $h=0$, the above action corresponds to the integrable theory of the thermal deformation, that has only one particle excitation and an elastic $S$-matrix equal to $S = -1$. On the contrary, when $T=T_c$ one recovers the integrable theory of the Ising in a magnetic field, whose spectrum consists of the $8$ massive particles, eq. (\ref{E8spectrum}). One may wonder how the spectrum evolves by moving the couplings, a question whose answer will be given in the next section.  
\item The second example is provided by the multi-frequency Sine-Gordon model, with Hamiltonian density 
\EQ
{\mathcal H}\,=\,\frac{1}{2}\left[(\partial_{t} \varphi)^2 + (\partial_x \varphi)^2\right] + \lambda_1   
\,\cos\beta \varphi + \lambda_2 \cos\alpha \varphi \,\,\,.
\EN 
When $\lambda_1 = 0$, this Hamiltonian gives rise to the integrable theory of the Sine-Gordon model with frequency $\alpha$.  In addition to the soliton states, such a theory has a number of neutral bound states given by\footnote{$[x]$ denotes the integer part of the real number $x$.}  $N_2 = \left[\frac{\pi}{\xi_{\alpha}}\right]$, 
where $\xi_{\alpha} = \frac{\alpha^2}{8}/(1 - \alpha^2/8\pi)$. Viceversa, if  $\lambda_2=0$, we have again a Sine-Gordon model but of frequency $\beta$ and a different number of neutral bound states, $N_1 = \left[\frac{\pi}{\xi_{\beta}}\right]$. When the ratio of $\alpha$ and $\beta$ is a rational number, the potential of the theory has an infinite number of periodic and degenerate vacua. On the contrary, when the ratio of the frequencies is an irrational number, the potential has only one vacuum that can always be placed at the origin. Also in this case, it is interesting to determine how the spectrum evolves by moving the two couplings.  
\end{itemize}
Notice that, to study the theories associated to an Hamiltonian as (\ref{multiple}), it is convenient to regard it as {\em a deformation of an integrable Hamiltonina} rather than as a multiple deformation of a conformal theory. By taking this point of view and grouping differently the terms, the Hamiltonian (\ref{multiple}) can be written as 
\EQ
 H = H^{i}_{\rm int}
+ \lambda_j \int \,dx \,\varphi_j(x) \,\,\, .
\label{group}
\EN
($i = 1,2$, $j\neq i$). There are several advantages in doing so. 
\begin{enumerate}
\item The first convenience is that the non-integrable theory can be analyzed starting from the basis of the Hilbert space provided by the particle excitations associated to the integrable model ${H}_{\rm int}^{i}$. Although the spectra of $H$ and $H_{int}^{i}$ would be different, the basis provided by the particles of the integrable model will be certainly more appropriate than the conformal one, as far as the infrared properties of the non-integrable model are concerned.
\item The second advantage consists of the exactly solvability of the integrable models, in particular, the possibility of computing exactly all the matrix elements (Form Factors) of local and non-local operators of such theories. Hence, in a complete analogy with ordinary Quantum Mechanics, one can conceive to set up a perturbative approach based on the Form Factors of the integrable models.  As we will see in the next Section, this perturbative approach will enable us to reach a remarkable series of predictions about the mass correction and the decay processes.  
\item When each deformation is individually integrable, there is the obvious freedom of using any of them as a starting point. By doing this choice we select a particular basis of the particles and bound states thereof. However, since the actual dynamics of the model should be insensitive to such a choice, there should be a series of mathematical identities that links one perturbative series to the other. 

\end{enumerate}

In the following we will discuss two different approaches to study 
the breaking of integrability: the Form Factor Perturbation Theory, the Semiclassical Method and the Truncated Conformal Space Approach. Each of them provides different information on non-integrable aspects.

\section{Form Factor Perturbation Theory}

For massive non--integrable field theories, a convenient perturbative scheme 
was originally proposed in \cite{DMS,DM} and called Form Factor Perturbation Theory 
(FFPT), since it is based on the knowledge of the exact Form Factors 
of the original integrable theory. Its generalization to massless case has been discussed in \cite{conmus}. In these papers it was shown that, even using just the first 
order correction of the FFPT, a great deal of information can be obtained,  
such as the evolution of their particle content, the variation of their masses 
and the change of the ground state energy. Whenever possible, universal ratios 
can be computed and successfully compared with their value obtained by 
other means, as the universal ratios 
relative to the decay of the particles with higher masses in the Ising model in 
a magnetic field, once the temperature is displayed away from the critical value \cite{Grinza}. For other and important aspects of the Ising model along non-integrable lines see the references \cite{McCoy,fonsecazam,rut,zamfon}.

\subsection{First Order Perturbation Theory}

Referring the reader to the original papers \cite{DMS,DM} for the theory of Form Factor Perturbation Theory, here we report the conclusions which can be reached by studying the first order correction to theory as the one of eq. (\ref{multiple}), where the integrable model is defined by the action (\ref{multiple}) with $\lambda_2 = 0$, whereas its perturbation is given by the relevant scalar operator $\varphi_2(x)$. Let  $x_1$ and $x_2$ be the scaling dimensions of the two operators $\varphi_1$ and $\varphi_2$.  The theory depends in this case on the two dimensionful couplings constants 
 $\lambda_1$ and $\lambda_2$ \footnote{For the sake of simplicity of notation, we assume that no other terms are generated by renormalisation effects, as it happens for the models that we will discuss later. The first order corrections do not depend on this assumption.}. Since $\lambda_1 \sim M^{2-x_1}$ and $\lambda_2 \sim M^{2-x_2}$ (where $M$ is a mass scale), we can decide to use $\lambda_1$ as dimensionful parameter of the theory and the dimensionless combination
\EQ
\chi \equiv \lambda_2 \,\lambda_1^{-\frac{2 - x_2}{2 - x_1}} 
\EN
as a label of the different Renormalization Group trajectories which originate from the fixed point at $\lambda_1 = \lambda_2 = 0$. For example, if $N(\chi)$ denotes the number of stable particles in the spectrum of the theory, their masses can be expressed as
\EQ
m_a(\lambda_1,\chi) 
,=\,{\mathcal C}_a(\chi)\,\lambda_1^{\frac{1}{2-x_1}}\,\,,\hspace{1cm}a=1,2,\ldots,N(\chi)\,\, ,
\label{ma}
\EN
where ${\mathcal C}_a(\chi)$ is an amplitude which characterises the whole trajectory. Similarly, the vacuum energy density can be written as
\EQ
{\mathcal E}_{\rm vac}(\lambda_1,\chi) = {\mathcal E}(\chi)\,
 \lambda_1^{\frac{2}{2-x_1}}\,\,\,.
\label{evac}
\EN
Dimensionless quantities, as for instance mass ratios, only depend on $\chi$ and therefore they do not vary along the trajectories of the Renormalization Group. 

Once the new interaction $\lambda_2 \int d^2x \,\varphi_2(x)$  is switched on in the action, the integrability of the unperturbed theory is generally lost and the S-matrix amplitudes become complicated quantities. Inelastic processes of particle production are no longer forbidden and, as a consequence, the analytic structure of the scattering amplitudes present additional cuts due to the higher thresholds. In particular, their expression is no longer factorized into the sequence of two-body scattering amplitudes and, even in elastic channels, the only surviving restriction on the final momenta comes from energy-momentum conservation. 

The knowledge of the matrix elements of the perturbing field $\varphi_2(x)$ ensures the possibility to compute perturbatively both the amplitudes of the inelastic processes and the corrections to the elastic ones. For instance, the first order corrections to the masses of the particles and to the vacuum energy density are given by 
\EQ
\delta M_{\bar{b}a}^2\, \simeq\, 2\lambda_2 \,F^{\varphi_2}_{\bar{b}a}(i\pi,0) \, \delta_{m_am_b}\,\,,
\label{dm}
\EN
\EQ
\delta {\mathcal E}_{\rm vac}\, \simeq\, \lambda_2\,\,\left[
{}_0\langle 0|\varphi_2|0\rangle_0\right]\,\,\,.
\label{de}
\EN
The best use of these formulas is to get rid of the explicit dependence on the normalisation of the perturbing operator by defining universal quantities, as ratios of the mass shifts. Hence, under the validity of the linear approximation, all the universal quantities of non-integrable field theories can be entirely expressed in terms of Form Factors of the integrable ones. A comparison of the theoretical predictions with their numerical determinations will be presented in the sequel, when we discuss the class of universality of the Ising model. Let's discuss now the important link which exists between the confinement of the excitations and the non-local nature of the perturbing operator.  

\noindent 
{\bf Non-locality and Confinement} Let's consider in more details the mass correction given in eq. (\ref{dm}) 
where the the form factor of the operator $\varphi_2(x)$ is defined by the matrix element 
\EQ
F_{a\bar{a}}^{\varphi_2}(\theta) \equiv \langle 0|\varphi_2(0)|
a(\theta_1)\bar{a}(\theta_2)\rangle\,\,\,.
\label{aabar}
\EN
Let us recall that the 2-particle Form Factor of an integrable theory satisfies the equations  
\EQ 
F_{a\bar{a}}^{\mathcal O}(\theta) \,=\,S_{a\bar{a}}^{b\bar{b}}(\theta)
F_{\bar{b}b}^{\mathcal O}(-\theta)\,\,,
\label{uni}
\EN
\EQ
F_{a\bar{a}}^{\mathcal O}(\theta + 2i\pi)\,=\, e^{-2i\pi\gamma_{{\mathcal O},a}}
F_{\bar{a}a}^{\mathcal O}(-\theta)\,\,.
\label{cross}
\EN
In the second equation the explicit phase factor 
 $e^{-2i\pi \gamma_{{\mathcal O},a}}$ is inserted to take into account a possible semi-locality of the operator which interpolates the particles and the operator  ${\mathcal O}(x)$. If $\gamma \neq 0$, the 2-particle Form Factor presents a pole at 
 $\theta =\pm i\pi$, with the residue given by  
\EQ
-i\,{\mbox Res}_{\theta =\pm i\pi}F_{a\bar{a}}^{\mathcal O}(\theta) \,=\,
(1-e^{\mp 2i\pi\gamma_{{\mathcal O},a}})\langle 0|{\mathcal O}|0\rangle\,\,.
\label{pole}
\EN
Hence, according whether the perturbing field is local or non-local with respect to the asymptotic particles, there are two different scenarios. If the field that breaks integrability is a local operator, the mass correction of the particles is finite. Viceversa, if the perturbing field is non-local, the mass correction of the particles is divergent. The last case implies the confinement of the particles, that occurs as soon as the non-integrable perturbation is turned on. 

There are several ways to show the confinement phenomena. One consists of computing the propagator  $\langle A(p) A(-p)\rangle $ of the particle $A$ in the perturbed theory. At the tree level approximation, this consists of a geometrical series that can be explicitly summed and for the propagator of the perturbed theory we have 
\EQ
\langle A(p) A(-p) \rangle \simeq \frac{1}{p^2 - m^2 - \delta m^2} \,\,\,.
\EN 
If $\delta m^2 = \infty$, the propagator obviously vanishes, i.e. the particle disappears from the spectrum. A more intuitive explanation of the confinement phenomenon comes from  the analysis of the Ising model, discussed below. 

Let's close this section by mentioning that, applied to the double Sine--Gordon model \cite{DM}, the FFPT has helped to clarify the rich dynamics of this non--integrable model. In particular, in relating the confinement of the kinks in the deformed theory 
to the non--locality properties of exponential operators with respect to the the kinks and in predicting the existence of a Ising--like phase transition for particular ratios of the two 
frequencies -- results which were later confirmed by a numerical 
study \cite{takacs}. The FFPT has been also used to study the spectrum of the 
$O(3)$ non-linear sigma model with a topological $\theta$ term, by varying $\theta$ 
\cite{CMPRL,conmus}, a model closely related to anti-ferromagnetic spin chains \cite{Haldane,affleck.lectures,ah}. 
 
\subsection{The scaling Region of the Ising Model}

\begin{figure}[t]
\centerline{\scalebox{0.5}
{\includegraphics{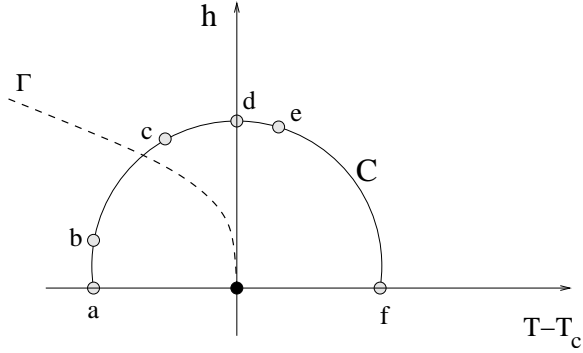}}} 
\caption{{\em Interpolation curve in the plane $(\tau,h)$ between the low and high-temperature phases of the Ising model.  $\Gamma$ is a Renormalization Group trajectory, identified by the dimensionless parameter $\chi$.}}
\label{curvaIsing}
\end{figure}

It is interesting to study the evolution of the mass spectrum of the Ising model by moving its couplings along the path $C$ shown in Fig. \ref{curvaIsing} in the plane  $(\tau,h)$ \cite{DMS}: this curve starts from the low temperature phase of the model and ends at its high-temperature phase, here represented by the points (a) and (f) respectively. 
The action of the model is given by 
\EQ
{\mathcal A} = {\mathcal A}_{CFT} + \tau \int d^2x\,
\varepsilon(x) + h \int d^2x\,\sigma(x)\,\, ,
\label{isingfield}
\EN
and gives rise to a massive integrable model when one of the two coupling constants is switched off. The action (\ref{isingfield}) defines a family of field theories  identified by $\chi\equiv \tau|h|^{-8/15}\in(-\infty,+\infty)$, a dimensionless RG invariant quantity. The spectrum of the theory changes in a significant way by moving $\chi$. 

In the low-temperature phase (corresponding to $\chi = -\infty$ and to the point (a) of the curve $C$), the model has two degenerate vacua and therefore its excitations consists of the topological kink and anti-kink that interpolated between the two ground states. Along the magnetic axes ($\chi=0$, corresponding to the point (d) of the curve $C$), the spectrum of the model consists instead of $8$ particles, with different masses. Finally, in the high-temperature phase 
(i.e. $\chi = + \infty$), the system has a unique vacuum and only a massive excitation above it. Let's see how this scenario can be recovered by the Form Factor Perturbation Theory. 

\begin{figure}[b]
\centerline{\scalebox{0.3}
{\includegraphics{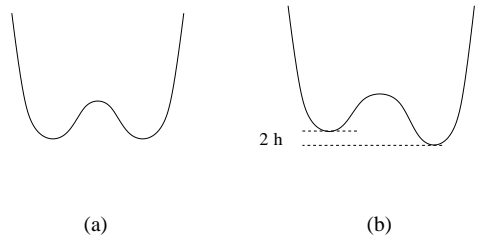}}} 
\caption{{\em Effecttive potential in the low-temperature phase (a) and in the presence of an infinitesimal magnetic field (b). In the last case, the two minima are no longer degenerate and the kink/antikink disappear from the spectrum of the asymptotic states.}}
\label{lift}
\end{figure}

Let's start our analysis from the point (a), where the massive excitations are the kink/anti-kinks that interpolate between the two degenerate vacua. By switching on the magnetic field, the model moves to the point (b) of the curve $C$. The form factor of the perturbing field $\sigma$ on the two-particle kink/anti-kink state is given by 
\[
F^{\sigma}(\theta_1-\theta_2) \,=\,\langle 0 | \sigma(0) | A(\theta_1) A(\theta_2)\rangle 
\,=\,\tanh\frac{\theta_1-\theta_2}{2}\,\,\,.
\] 
Therefore eqn. (\ref{dm}) leads to an infinite correction to the mass of the kinks, i.e. the kinks get confined as soon as the magnetic field is switched on. Looking at the effective potential of the theory, it is not difficult to see that this is the correct conclusion: no matter how small the magnetic field may be, it lifts the degeneracy of the two vacua, as shown in Fig. \ref{lift}, and consequently there is no longer the possibility of having topological configurations. 

\noindent

\begin{figure}[b]
\centerline{\scalebox{0.4}
{\includegraphics{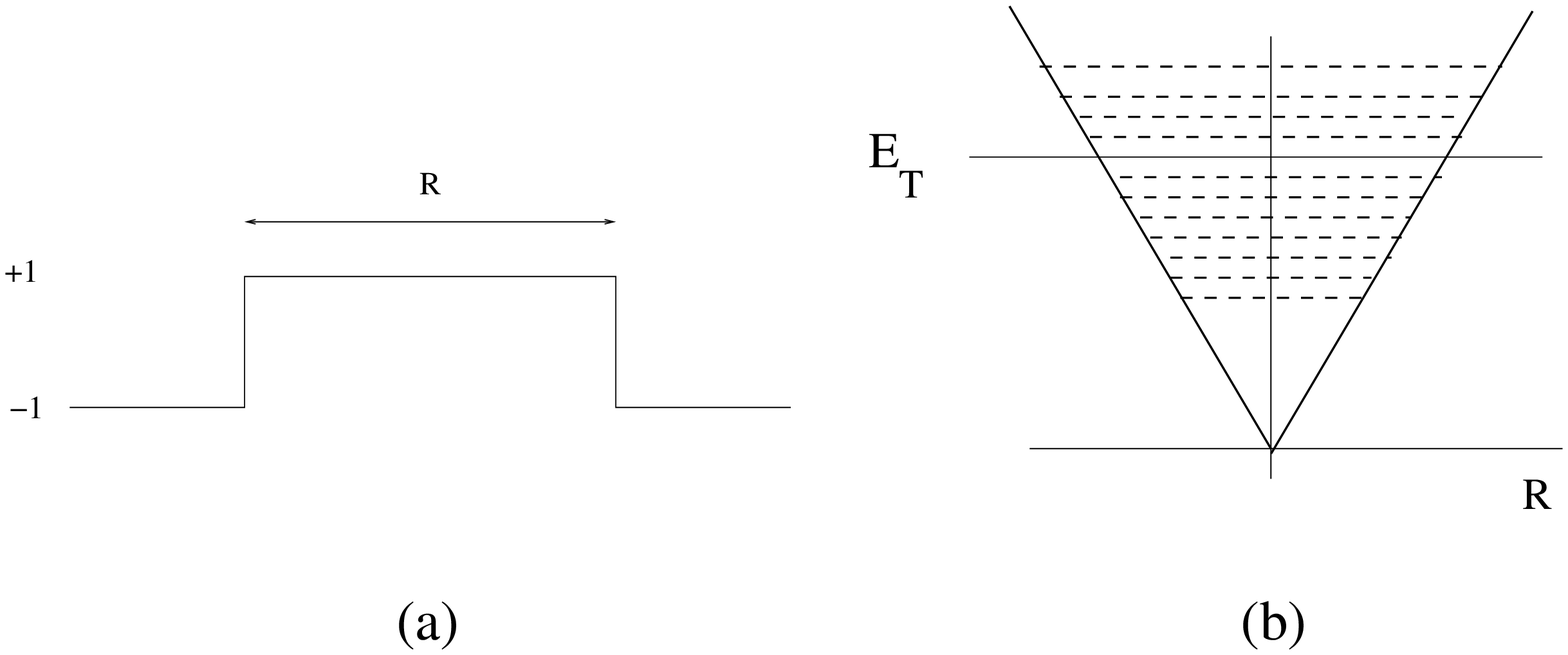}}} 
\caption{{\em (a): kink-antikink state separated by a distance $R$; (b) kink-antikink potential in the presence of a magnetic field $h$ and its bound states. The  stable bound states are identified by the condition $E_n < E_T$.}}
\label{kinkanti}
\end{figure}

Consider now the effect of the magnetic field on a state made of a kink and an antikink separated by a distance $R$. When the magnetic field is absent, the energy of this state is essentially equal to $2 M$, i.e. the sum of the masses of the kink and the antikink. 
The energy of this state depends very weakly on the distance $R$ because this field configuration takes values on the zeros of the effective potential and, no matter how large the distance $R$ could be, there is no change in the energy of this state. This situation changes by switching on the magnetic field since, in this case, at every point of the space there is an energy gap equal to $2 h$ and the energy $U$ of this state becomes a linear function of $R$,  $U(R) = 2 M + 2 h R$. This attractive interaction between the kink and the antikink gives rise to a discrete spectrum of bound states. Regarding the kinks as very massive and quasi-static,  the energy of the bound states can be obtained by solving the quantum mechanical problem of the bound states for a linear potential, well-known in Quantum Mechanics.  The result is simply  \cite{McCoy}
\EQ
E_k \, \equiv \, m_k\,=\,(2 + h^{2/3} \gamma_k^{2/3}) M  \,\, ,
\label{chain}
\EN
where  $\gamma_k$ are the positive roots of the equation 
\[
{\mathcal J}(\gamma_k)\,=\,J_{\frac{1}{3}} \left(\frac{1}{3} \gamma_k\right) + 
J_{-\frac{1}{3}} \left(\frac{1}{3} \gamma_k\right) = 0 \,\, ,
\]
($J_{\nu}(x)$ is the Bessel function of order $\nu$). The structure of the bound states is shown in Fig. \ref{kinkanti}. Obviously not all these states are stable: the stable ones are identified by the condition  $m_n < 2 m_1$, while all particles with a mass higher than the threshold $2 m_1$ decay into particles of lower masses.
When $\chi$ increases, i.e. when the system moves clockwise along the curve $C$ of Fig.  
\ref{curvaIsing}, the number of bound states monotonically decreases. At the point (d), there are the $8$ stable particles of the Zamolodchikov's solution of the Ising model in a magnetic field.   

It is worth stressing that in this case the five particles with mass higher than the threshold are stable just for the integrability of the model. Moving away from the magnetic axes by means of the operator $\varepsilon(x)$, the first three particles change the value of their masses, while the remaining five particles decay into the low-energy channels. To estimate both effects, we need the Form Factors of the energy operator but in the integrable theory of the Ising model in a magnetic field, determined in 
\cite{DelSim}. Here we simply report their expressions  
\begin{eqnarray*}
&& \langle 0|\varepsilon(0)|0\rangle\, = \,m_1\,\,,\nonumber\\
&& F_{11}^\varepsilon(i\pi)\, =\, \langle
0|\varepsilon(0)|A_1(\theta + i\pi) A_1(\theta )\rangle\,=\, -17.8933..m_1\,\,,\nonumber\\
&& F_{22}^\varepsilon(i\pi)\, =\, \langle
0|\varepsilon(0)|A_2(\theta + i\pi) A_2(\theta)\rangle \,=\, -24.9467..m_1\,\,,\nonumber\\
&& F_{33}^\varepsilon(i\pi)\,=\,\langle
0|\varepsilon(0)|A_3(\theta + i\pi) A_3(\theta)\rangle\,=\, -53.6799..m_1\,\,,\nonumber\\
\label{energyff}
\end{eqnarray*}
The first equation may be regarded as the normalization condition of the energy operator $\varepsilon(x)$. The corrections of the universal ratios are given by 
\begin{eqnarray}
&& \frac{\delta{\mathcal E}_{vac}}{\delta m_1}=
\frac{\langle 0|\varepsilon|0\rangle}
{F_{11}^\varepsilon(i\pi)}\,m_1^0\, = \, -0.0558..m_1^0\,\,,\nonumber\\
&& \frac{\delta m_2}{\delta m_1} \,=\, 
\frac{F_{22}^\varepsilon(i\pi)}{F_{11}^\varepsilon(i\pi)}
\,\frac{m_1^0}{m_2^0} \,= \, 0.8616..\,\,,\label{th}\\
&& \frac{\delta m_3}{\delta m_1} \, = \, \frac{F_{33}^\varepsilon(i\pi)}
{F_{11}^\varepsilon(i\pi)}\,\frac{m_1^0}{m_3^0} \,= \,1.5082..\,\,\,.\nonumber
\end{eqnarray}
In turn, these quantities can be independently determined by a numerical solution of the model \cite{DMS} and the values determined in this way are 
\begin{eqnarray}
&& \frac{\delta{\mathcal E}_{vac}}{\delta m_1} \simeq -0.05\,m_1^0\,\,,\nonumber\\
&& \frac{\delta m_2}{\delta m_1} \simeq 0.87\,\,, \label{measure}\\
&& \frac{\delta m_3}{\delta m_1} \simeq 1.50\,\,\,.\nonumber
\end{eqnarray}
As it can be seen from the expressions above, there is a satisfactory agreement between the theoretical and the numerical estimates. 

\begin{figure}[t]
\centerline{\scalebox{0.4}
{\includegraphics{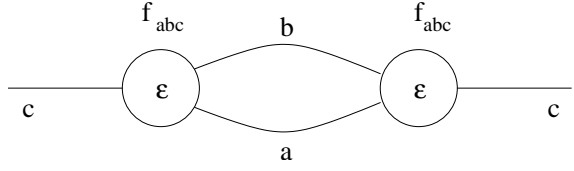}}} 
\caption{{\em Perturbative diagram at the second order in $\tau$ relative to the imaginary part of the mass of the particle $c$. The intermediate particles  $a$ and $b$ satisfy the on-shell conditions $p^2_a = m^2_a$ e $p^2_b = m^2_b$. When $c > 5$, there are additional diagrams with more intermediate particles.}}
\label{deltaM}
\end{figure}

Breaking the integrability of the Ising model in a magnetic field has a more dramatic effect on the five particles with a mass above threshold. Their stability is only due to integrability and, in its absence, they decay. These decay processes were studied in 
\cite{Grinza}. In the perturbative approach, the decay processes are associated to the presence of a negative imaginary part in the mass that is a second order perturbative effect in $\tau$, as shown in Fig. \ref{deltaM}. 

\[
\mbox{Im}\,m_c^2 \, = \, -\sum_{a\leq b\,,\,m_a + m_b\leq m_c} m_c\,
\Gamma_{c\to ab}\simeq -\tau^2\sum_{a\leq b\,,\,m_a+m_b\leq m_c}
2^{1-\delta_{ab}}\frac{|f_{cab}|^2}
{m_cm_a\left|\sinh\theta^{(cab)}_a\right|}\,,
\label{im}
\]
where $\Gamma_{c \to ab}$ is the decay amplitude of the particle $A_c$ into the two particles $A_ a \,A_b$, whereas 
\[
f_{cab}\,=\,\left.F^\varepsilon_{cab}(i\pi,\theta^{(cab)}_a,\theta^{(cab)}_b)
\right|_{\tau=0}\,. 
\]
The rapidities  $\theta^{(cab)}_a$ and $\theta^{(cab)}_b$ are fixed by the conservation of the energy and the momentum in the decay process $A_c\rightarrow A_aA_b$ in the rest frame of the particle $A_c$. In the above equations all masses are the unperturbed  values at $\tau=0$. When $c>5$ the sum must be completed including the contribution of the decay channels with more than two particles in the final state. Once the decay amplitudes $\Gamma_{c\rightarrow a b}$ are known, one can determine the lifetime $t_c$ of the unstable particle  $A_c$ given by 
\EQ
t_c=\frac{1}{\Gamma_c}\,,\hspace{1cm}
\Gamma_c=\sum_{a\leq b} \Gamma_{c\to ab}\,\,.
\EN
For the Ising model, the relevant matrix elements are  
\begin{eqnarray*}
&& |f_{411}|=(36.73044..)\left|\langle\varepsilon\rangle\right|_{\tau=0}
\label{f411}\\
&& |f_{511}|=(19.16275..)\left|\langle\varepsilon\rangle\right|_{\tau=0}\\
&& |f_{512}|=(11.2183..)\left|\langle\varepsilon\rangle\right|_{\tau=0}
\,\,.
\label{f512}
\end{eqnarray*}
where the normalization of the operator $\varepsilon$ is fixed by its vacuum expectation value 
\[
\langle\varepsilon\rangle_{\tau=0}=(2.00314..)\,|h|^{8/15}\,\,.
\]
The imaginary part of the mass of the first two particles, which are over threshold, is given by 
\begin{eqnarray*}
&& \mbox{Im}\,m_4^2\simeq(-840.172..)\left(\frac{\tau
\langle\varepsilon\rangle_{\tau=0}}{m_1}\right)^2=(-173.747..)\,\tau^2\\
&& \mbox{Im}\,m_5^2\simeq(-240.918..)\left(\frac{\tau
\langle\varepsilon\rangle_{\tau=0}}{m_1}\right)^2=(-49.8217..)\,\tau^2\,\,.
\end{eqnarray*}
The ratio of their lifetime is universal 
\EQ
\lim_{\tau\to 0}\frac{t_4}{t_5}\,=\,\lim_{\tau\to 0}
\frac{m_4\,\mbox{Im}\,m_5^2}{m_5\,\mbox{Im}\,m_4^2}\,=\,0.23326..
\label{liferatio}
\EN
While the particle $A_4$ can only decay into  $A_1A_1$, the particle $A_5$ can also decay into the channel $A_1A_2$. The ratios of the amplitudes of these decays 
\[
b_{c\to ab}\,=\,\frac{m_c|_{\tau=0}\,\Gamma_{c\to ab}}{|\mbox{Im}\,m_c^2|}
\]
are given by 
\[
\lim_{\tau\to 0}b_{5\to 11}\,=\,0.47364..\,,\hspace{1cm}
\lim_{\tau\to 0}b_{5\to 12}\,=\,0.52635..\,\,.
\]
Notice that eq.\,(\ref{liferatio}) predicts that the lifetime of the particle $A_5$ is almost four time longer than the lifetime of the particle $A_4$. This paradoxical result, in contradiction with the intuitive idea that a heavy particle should decay faster than a light one, finds its explanation once again in the peculiar behavior of the phase space in two dimensions. For the decay process $A_c\to A_aA_b$ the phase space in $d$-dimension is given by 
\EQ
\int\frac{d^{d-1}\vec{p}_a}{p^0_a}\,\,\frac{d^{d-1}\vec{p}_b}{p^0_b}\,\,
\delta^d(p_a-p_b)\sim\frac{p^{d-3}}{m_c}\,,
\EN
where $p=|\vec{p}_a|=|\vec{p}_b|$ is the value in the rest frame. For fixed decay products, $p$ grows with $m_c$: in $d=2$, this term joins the factor $m_c$ in the denominator and leads to a suppression of the phase space. In the Ising model, eqns.  (\ref{f411})--(\ref{f512}) show that this suppression is further enhanced by the dynamics (i.e. by the vaues of the matrix elements) in a way that is not compensated by the additional decay channels.  

If we keep moving along the curve $C$, one firstly meets a value $\chi_1$ at which the mass of the particle $A_3$ becomes larger than $2 m_1$ and, later on, a second value $\chi_2$ at which also the mass of the particle $A_2$ becomes larger than $2 m_1$. When $\chi > \chi_2$ the spectrum of the stable particles of the theory consists of only one excitation. In the limit $\chi \rightarrow + \infty$, this is nothing but the particle of the integrable theory of the high-temperature phase of the model. 

\section{Semiclassical methods}\label{intro}

\begin{figure}[b]
\centerline{\scalebox{0.5}
{\includegraphics{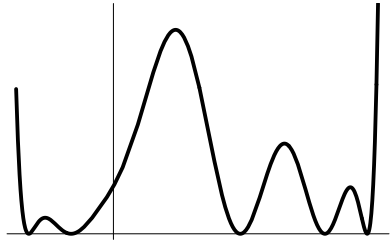}}} 
\caption{{\em Potential $U(\varphi)$ of a quantum field theory with kink 
excitations.}}
\label{potential}
\end{figure}

Non-integrable quantum field theories with topological excitations (kink-like) can be also studied by semiclassical methods. Such theories are
described by a scalar real field $\varphi(x)$, with a Lagrangian density 
\EQ
{\cal L} \,=\,\frac{1}{2} (\partial_{\mu} \varphi)^2 - U(\varphi) \,\,\,, 
\label{Lagrangian}
\EN 
where the potential $U(\varphi)$ possesses several degenerate minima at 
$\varphi_a^{(0)}$ ($a =1,2,\ldots,n$), as the one shown in Figure \ref{potential}. These 
minima correspond to the different vacua $\mid \,a\,\rangle$ of 
the associate quantum field theory. 


\vspace{1mm}

\begin{figure}[t]
\centerline{\scalebox{0.3}
{\includegraphics{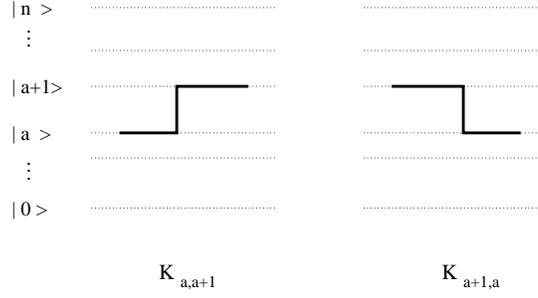}}} 
\caption{{\em Kink and antikink configurations.}}
\label{step}
\end{figure}

\begin{figure}[b]
\centerline{\scalebox{0.3}
{\includegraphics{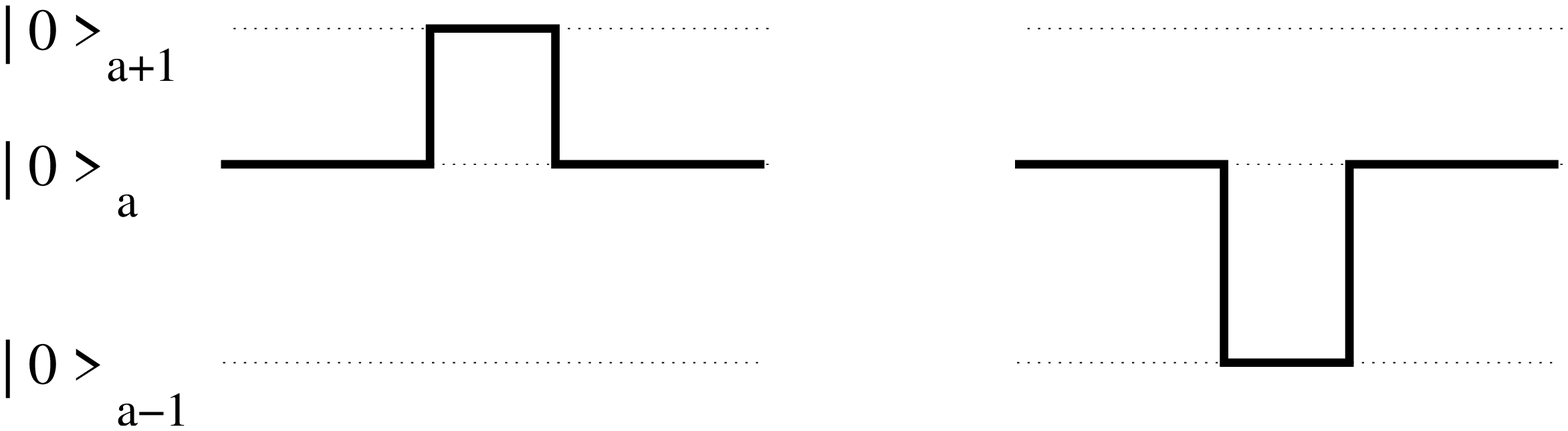}}} 
\caption{{\em Kink-antikink configurations which may give rise to a bound state nearby the vacuum 
$\mid 0\,\rangle_a$.}}
\label{tooth}
\end{figure}

\noindent
{\bf Excitations of the system}. The basic excitations of this kind of models are kinks and anti-kinks, i.e. topological configurations which interpolate between two neighbouring vacua. Semiclassically they correspond to the static solutions of the equation of motion, i.e. 
\EQ
\partial^2_x \,\varphi(x) \,=\,U'[\varphi(x)] \,\,\,, 
\label{static}
\EN
with boundary conditions $\varphi(-\infty) = \varphi_a^{(0)}$ and 
$\varphi(+\infty)= \varphi_{b}^{(0)}$, where $b = a \pm 1$. 
Denoting by $\varphi_{ab}(x)$ the solutions of this equation, their 
classical energy density is given by 
\EQ
\epsilon_{ab}(x) \,=\,\frac{1}{2} \left(\frac{d\varphi_{ab}}{d x}\right)^2 + U(\varphi_{ab}(x))  \,\,\,,
\EN 
and its integral provides the classical expression of the kink masses 
\EQ
M_{ab} \,=\,\int_{-\infty}^{\infty} \epsilon_{ab}(x) \,\,\,.
\label{integralmass}
\EN   
It is easy to show that the classical masses of the kinks $\varphi_{ab}(x)$ are 
simply proportional to the heights of the potential between the two minima 
$\varphi_a^{(0)}$ and $\varphi_b^{(0)}$.  Once put in motion by a a Lorentz transformation, i.e. $\varphi_{ab}(x) \rightarrow \varphi_{ab}\left[(x \pm v t)/\sqrt{1-v^2}\right]$, these configurations describe in the quantum theory the kink states $\mid K_{ab}(\theta)\,\rangle$, where $a$ and $b$ are the indices of the initial and final vacuum, respectively. The quantity $\theta$ is the rapidity variable which parameterises the relativistic dispersion relation of these excitations, i.e. 
\EQ
E = M_{ab}\,\cosh\theta
\,\,\,\,\,\,\,
,
\,\,\,\,\,\,\,
P = M_{ab} \,\sinh\theta
\,\,\,.
\label{rapidity}
\EN 
Conventionally $\mid K_{a,a+1}(\theta) \,\rangle$ denotes the {\it kink} between the pair 
of vacua $\left\{\mid a\,\rangle ,\mid a+1\,\rangle\right\}$ while $\mid K_{a+1,a}\,\rangle$ 
is the corresponding {\it anti-kink}. For the kink configurations one can adopt 
the simplified graphical form shown in Figure \ref{step}.
The multi-particle states are given by a string of these excitations, with the adjacency 
condition of the consecutive indices for the continuity of the field configuration
\EQ
\mid K_{a_1,a_2}(\theta_1) \,K_{a_2,a_3}(\theta_2)\,K_{a_3,a_4}(\theta_3) \ldots \rangle 
\,\,\,\,\,\,\,\,
,
\,\,\,\,\,\,\,\, (a_{i+1} = a_i \pm 1)
\EN 
In addition to the kinks, in the quantum theory there may exist other excitations, i.e. 
ordinary scalar particles (breathers). These are the neutral excitations $\mid B_c(\theta)\,\rangle_a$ ($c=1,2,\ldots$) around each of the vacua $\mid a\,\rangle$. For a theory based on a Lagrangian of a single real field, these states are all non-degenerate: in fact, there are no extra quantities which commute with the Hamiltonian and that can give rise to a multiplicity of them. The neutral particles must be identified as the bound states of the kink-antikink configurations that start and end at the same vacuum $\mid a\,\rangle$, i.e. $\mid K_{ab}(\theta_1) \,K_{ba}(\theta_2)\,\rangle$, with the ``tooth'' shapes shown in Figure\,\,\ref{tooth}. 
If such two-kink states have a pole at an imaginary value $i \,u_{a b}^c$ within the physical strip $0 < {\rm Im}\, \theta < \pi$ of their rapidity difference $\theta = \theta_1 - \theta_2$, then their bound states are defined through the factorization formula which holds in the vicinity of this singularity 
\EQ
\mid K_{ab}(\theta_1) \,K_{ba}(\theta_2) \,\rangle \,\simeq \,i\,\frac{g_{ab}^c}{\theta - i u_{ab}^c}
\,\mid B_c\,\rangle_a \,\,\,.
\label{factorization}
\EN 
In this expression $g_{ab}^c$ is the on-shell 3-particle coupling between the kinks and the neutral particle. Moreover, the mass of the bound states is simply obtained by substituing the resonance value $i \,u_{ab}^c$ within the expression of the Mandelstam variable $s$ of the two-kink channel 
\EQ
s = 4 M^2_{ab} \,\cosh^2\frac{\theta}{2} 
\,\,\,\,\,\,
\longrightarrow 
\,\,\,\,\,\, 
m_c \,=\,2 M_{ab} \cos\frac{u_{ab}^c}{2} \,\,\,.
\label{massboundstate}
\EN 
Concerning the vacuum states, in the infinite volume their classical degeneracy is removed by selecting one of them, say  $\mid k \,\rangle$, out of the $n$ available. This happens through a "spontaneously symmetry breaking mechanism", even though there may be no internal symmetry to break at all, as for the potential of Figure \ref{potential}. 
In the absence of a symmetry which connects the various vacua, the excitations above them may be very different and it is interesting to address the question of their  
and the value of their masses. This problem can be solved through the application of a  remarkably simple formula due to Goldstone-Jackiw \cite{GJ}: this applies in the semiclassical approximation, i.e. when the coupling constant goes to zero and 
the mass of the kinks becomes correspondingly very large with respect to any other mass scale. In its refined version, given in \cite{FFvolume}, this formula reads as follows 
\EQ
f_{ab}^{\varphi}(\theta) \,=\,\langle K_{ab}(\theta_1) \,\mid \varphi(0) \,\mid \, 
K_{ab}(\theta_2) \rangle 
\,\simeq \,\int_{-\infty}^{\infty} dx \,e^{i M_{ab} \,\theta\,x} \,\,
\varphi_{ab}(x) \,\,\,,
\label{remarkable1}
\EN  
where $\theta = \theta_1 - \theta_2$. Substituting in this formula $\theta \rightarrow i \pi - \theta$, the corresponding expression may be interpreted as the following Form Factor 
\EQ
F_{ab}^{\varphi}(\theta) \,=\, f(i \pi - \theta) \,=\,\langle a \,\mid \varphi(0) \,\mid \,
K_{ab}(\theta_1) \, K_{ba}(\theta_2) \rangle \,\,\,,
\label{remarkable2}
\EN    
where appears the neutral kink states around the vacuum $\mid a \rangle$ 
of interest. Eq.\,(\ref{remarkable1}) deserves several comments. 

\begin{enumerate}
\item 
The appealing aspect of the formula (\ref{remarkable1}) stays in the relation between the Fourier transform of the {\em classical} configuration of the kink, -- i.e. the solution 
$\varphi_{ab}(x)$ of the differential equation (\ref{static}) -- to the {\em quantum} 
matrix element of the field $\varphi(0)$ between the vacuum $\mid a\,\rangle$ and the 
2-particle kink state $\mid K_{ab}(\theta_1) \,K_{ba}(\theta_2)\,\rangle$. 

Given the solution of eq.\,(\ref{static}) and its Fourier transform, the poles of $F_{ab}(\theta)$ within the physical strip of $\theta$ identify the neutral bound states which couple to $\varphi$. The mass of the neutral particles can be extracted by using eq.\,(\ref{massboundstate}), while the on-shell 3-particle coupling $g_{ab}^c$ can be obtained from the residue at these poles (Figura \ref{residuef})  
\EQ
\lim_{\theta \rightarrow i \,u_{ab}^c} (\theta - i u_{ab}^c)\, F_{ab}(\theta)
\,=\,i \,g_{ab}^c \,\,\langle a \,\mid \varphi(0) \,\mid \,B_c \,\rangle \,\,\,.
\label{residue}
\EN  

\begin{figure}[t]
\centerline{\scalebox{0.4}
{\includegraphics{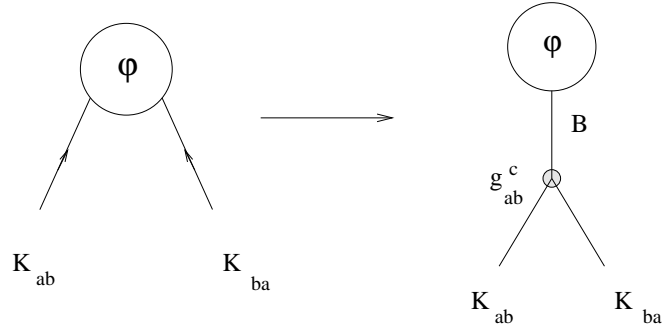}}} 
\caption{{\em Residue equation for the matrix element on the kink states.}}
\label{residuef}
\end{figure}

\item For a generic theory, the classical kink configuration $\varphi_{ab}(x)$ is 
not related in a simple way to the anti-kink configuration $\varphi_{ba}(x)$.  This is the reason why neighbouring vacua may have a different spectrum of neutral excitations, as shown in the examples discussed below.    

\item This procedure for extracting the bound states masses permits in many cases 
to avoid the semiclassical quantization of the breather solutions \cite{DHN}, making their derivation much simpler. The reason is that, the classical breather configurations depend also on time and have, in general, a more complicated mathematical structure than the kink ones. Moreover, it can be shown that in non--integrable theories 
these configurations do not exist as exact solutions of the partial differential equations of the field theory. On the contrary, to apply eq.\,(\ref{remarkable1}), one simply needs the solution of the {\em ordinary} differential equation (\ref{static}). It is worth notice that, to locate the 
poles of $f_{ab}^{\varphi}(\theta)$, one has simply to look at the exponential behavior of the classical solutions at $x \rightarrow \pm \infty$, as discussed below. 
\end{enumerate}

Below we will discuss a class of theories with only two vacua, which can be either symmetric or asymmetric ones. A complete analysis of other potentials can be found in the original paper \cite{kink}.  

\subsection{Symmetric wells}\label{phi4section}

A prototype example of a potential with two symmetric wells is the $\varphi^4$ theory in its broken phase. The potential is given in this case by 
\EQ
U(\varphi) \,=\,\frac{\lambda}{4} \left(\varphi^2 - \frac{m^2}{\lambda}\right)^2 \,\,\,.
\EN 
Let us denote with $\mid \pm 1 \,\rangle$ the vacua corresponding to the classical 
minima $\varphi_{\pm}^{(0)} \,=\,\pm \frac{m}{\sqrt{\lambda}}$. By expanding around them, $\varphi \,=\,\varphi_{\pm}^{(0)} + \eta$, we have 
\EQ
U(\varphi_{\pm}^{(0)} + \eta) \,=\, m^2 \,\eta^2 \pm m \sqrt{\lambda}\,\eta^3 + \frac{\lambda}{4}
\eta^4 \,\,\,.
\label{potentialphi4}
\EN 
Hence, perturbation theory predicts the existence of a neutral particle for each of the two vacua, with a bare mass given by $m_b = \sqrt{2} m$, irrespectively of the value of the coupling $\lambda$. Let's see, instead, what is the result of the semiclassical analysis. The kink solutions are given in this case by
\EQ
\varphi_{-a,a}(x) \,=\,a \,\frac{m}{\sqrt{\lambda}} \,\tanh\left[\frac{m x}{\sqrt{2}}\right]
\,\,\,\,\,\,\,
,
\,\,\,\,\,\,\, a = \pm 1 
\label{kinksolphi4}
\EN 
and their classical mass is 
\EQ
M_0\,=\,\int_{-\infty}^{\infty} \epsilon(x) \,dx \,=\,\frac{2 \sqrt{2}}{3} \,
\frac{m^3}{\lambda}  
\,\,\,.
\EN 
The value of the potential at the origin, which gives the height of the barrier between the two vacua, can be expressed as 
\EQ
U(0) \,=\,\frac{3 m}{8 \sqrt{2}} \,M_0 \,\,\,,
\EN 
and is proportional to the classical mass of the kink. If we take into account the contribution of the small oscillations around the classical static configurations, the kink mass gets corrected as \cite{DHN}
\EQ
M \,=\,\frac{2 \sqrt{2}}{3} \,\frac{m^3}{\lambda} - m 
\left(\frac{3}{\pi \sqrt{2}} - \frac{1}{2 \sqrt{6}}\right) 
+ {\cal O}(\lambda) \,\,\,.
\label{mass1phi4}
\EN 
Defining  
\[
c =  \left(\frac{3}{2\pi} - \frac{1}{4 \sqrt{3}}\right) > 0 \,\,\,,
\]
and the adimensional quantities
\EQ
g = \frac{3 \lambda}{2 \pi m^2}
\,\,\,\,\,\,\,\,
;
\,\,\,\,\,\,\,\,
\xi \,=\,\frac{g}{1 - \pi c g} \,\,\,, 
\label{definitiong}
\EN 
the mass of the kink can be expressed as 
\EQ
M \,=\,\frac{\sqrt{2} m}{\pi \,\xi}\,=\,\frac{m_b}{\pi \,\xi}\,\,\,.
\label{newmassphi4}
\EN
Since the kink and the anti-kink solutions are equal functions (up to a sign), their Fourier transforms have the same poles and therefore the spectrum of the neutral particles will be the same on both vacua, in agreement with the $Z_2$ symmetry of the model. We have  
\EQ
f_{-a,a}(\theta) \,= \, \int_{-\infty}^{\infty} 
dx \,e^{i M  \theta\,x} \varphi_{-a,a}(x)  \nonumber \, =\, 
 i\,a \sqrt{\frac{2}{\lambda}}\,\frac{1}{\sinh\left(\frac{\pi M}{\sqrt{2} m} \theta\right)}\,\,\,.
\EN
By making now the analitical continuation $\theta \rightarrow i \pi - \theta$ and using the above definitions (\ref{definitiong}), we arrive to   
\EQ
F_{-a,a}(\theta) \,= \,
\langle a\,\mid \varphi(0)\,\mid K_{-a,a}(\theta_1) K_{a,-a}(\theta_2) \rangle 
\,\propto \,\,  
\frac{1}{\sinh\left(\frac{(i \pi - \theta)}{\xi}\right)} \,\,\,.
\label{FFphi4}
\EN 
The poles of the above expression are located at 
\begin{equation}
\theta_{n}\,=\,i \pi \left(1 - \xi \,n\right)
\,\,\,\,\,\,\,
,
\,\,\,\,\,\,\,
n = 0,\pm 1,\pm 2,\ldots
\label{polesphi4}
\end{equation}
and, if 
\EQ
\xi \geq 1 \,\,\,,
\label{conditionphi4}
\EN 
none of them is in the physical strip $0 < {\rm Im}\,\theta < \pi$. Consequently, in the 
range of the coupling constant  
\EQ
\frac{\lambda}{m^2} \geq \frac{2 \pi}{3} \,\frac{1}{1 + \pi c} \,=\,1.02338...
\label{criticalg}
\EN 
the theory does not have any neutral bound states, neither on the vacuum to the right nor on the one to the left. Viceversa, if $\xi < 1$, there are $n = \left[\frac{1}{\xi}\right]$ neutral bound states, where $[ x ]$ denote the integer part of the number $x$. Their semiclassical masses are given by 
\begin{equation}
m_{b}^{(n)} \,=\, 2 M\,\sin\left[n\frac{\pi \xi}{2}\,\right]\,=\,
n\,\,m_b\left[1 - \frac{3}{32}\,\frac{\lambda^2}{m^4} 
\,n^{2} +...\right]\,.
\label{massphi4}
\end{equation}
Note that the leading term is given by multiples of the mass of the elementary boson $\mid B_1\rangle$.  Therefore the $n$-th breather may be considered as a loosely bound state of $n$ of it, with the binding energy provided by the remaining terms of the above expansion. But, for the non-integrability of the theory, all particles with mass $m_n > 2 m_1$ will eventually decay. It is easy to see that, if there are at most two particles in the spectrum, it is always valid the inequality $m_2 < 2 m_1$. However, 
if $\xi < \frac{1}{3}$, for the higher particles one always has 
\EQ
m_k > 2 m_1
\,\,\,\,\,\,\,
,
\,\,\,\,\,\,\,
{\makebox for}\,\, k=3,4,\ldots n \,\,\,.
\EN 
According to the semiclassical analysis, the spectrum of neutral particles of $\varphi^4$ theory is then as follows: (i) if $\xi > 1$, there are no neutral particles; (ii) if $\frac{1}{2} < \xi < 1$, there is one particle; (iii) if $\frac{1}{3} < \xi < \frac{1}{2}$ there are two particles; (iv) if $\xi < \frac{1}{3}$ there are $\left[\frac{1}{\xi}\right]$ particles, although only the first two are stable, because the others are resonances.  

\begin{figure}[t]
\centerline{\scalebox{0.3}
{\includegraphics{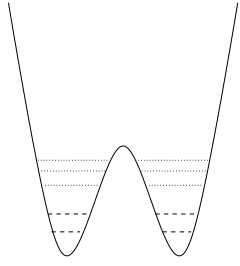}}} 
\caption{{\em Neutral bound states of $\varphi^4$ theory for $g < 1$. The lowest two lines are 
the stable particles whereas the higher lines are the resonances.}}
\label{doublewell}
\end{figure}

Let us now briefly mention some general features of the semiclassical methods, starting from an equivalent way to derive the Fourier transform of the kink solution. To simplify the notation, let's get rid of all possible constants and consider the Fourier transform of the derivative of the kink solution, expressed as 
\EQ
G(k) \,=\,\int_{-\infty}^{\infty} d x \,e^{i k x} \frac{1}{\cosh^2 x} \,\,\,. 
\EN 
We split the integral in two terms
\EQ
G(k) \,=\,\int_{-\infty}^0 dx \,e^{i k x} \,\frac{1}{\cosh^2 x} 
+ \int_0^{\infty} dx \,e^{i k x} \,\frac{1}{\cosh^2 x} 
\,\,\,,
\label{FT}
\EN 
and we use the following series expansion of the integrand, valid on the entire real axis (except the origin)  
\EQ
\frac{1}{\cosh^2 x} \,=\,4 \,\sum_{n=1}^{\infty} (-1)^{n+1} n \,e^{-2 n |x|} \,\,\,.
\EN 
Substituting this expression into (\ref{FT}) and computing each integral, we have 
\EQ
G(k) \,=\,4 i\sum_{n=1}^{\infty} (-1)^{n+1} n \left[-\frac{1}{i k + 2n} + \frac{1}{-i k + 2n }\right] 
\,\,\,. 
\EN 
Obviously it coincides with the exact result, $G(k) \,=\,\pi k/\sinh\frac{\pi}{2} k$, but this derivation permits to easily interpret the physical origin of each pole. In fact, changing $k$ to the original variable in the crossed channel, $k \rightarrow (i \pi - \theta)/\xi$, we see that the poles which determine the bound states at the vacuum $\mid a \rangle$ are only those relative to the exponential behaviour of the kink solution at $x \rightarrow -\infty$. This is precisely the point where the classical kink solution takes 
values on the vacuum $\mid a \rangle$. In the case of $\varphi^4$, the kink and the antikink are the same function (up to a minus sign) and therefore they have the same exponential approach at $x = -\infty$ at both vacua $\mid \pm 1 \rangle$. Mathematically speaking, this is the reason for the coincidence of the 
bound state spectrum on each of them: this does not necessarily happens in other cases, as we will see in the next section, for instance. 

The second comment concerns the behavior of the kink solution near the minima of the potential. In the case of $\varphi^4$, expressing the kink solution as 
\EQ
\varphi(x) \,=\,\frac{m}{\sqrt \lambda} \,\tanh\left[\frac{m \,x}{\sqrt 2}\right] \,=\,
\frac{m}{\sqrt \lambda}\,\,\frac{e^{\sqrt{2} \,x} -1}{e^{\sqrt{2}\,x} + 1} \,\,\,,
\EN 
and expanding around $x = -\infty$, we have 
\EQ
\varphi(t) \,=\,-\frac{m}{\sqrt \lambda}\,\left[1 - 2 t + 2 t^2 - 2 t^3 + \cdots 2 \,(-1)^n t^n \cdots \right] 
\,\,\,,
\EN 
where $t = \exp[\sqrt{2} x]$. Hence, all the sub-leading terms are exponential factors, with exponents which are multiple of the first one. Is this a general feature of the kink solutions of any theory? It can be proved that the answer is indeed positive \cite{kink}. 

The fact that the approach to the minimum of the kink solutions is always through multiples of the same exponential (when the curvature $\omega$ at the minimum is different from zero) implies that the Fourier transform of the kink solution has poles regularly spaced by $\xi_a \equiv \frac{\omega}{\pi M_{ab}}$ in the variable $\theta$. If the first of them is within the physical strip, the semiclassical mass spectrum derived from the formula (\ref{remarkable1}) near the vacuum $\mid a \,\rangle$ has therefore the universal form 
\EQ
m_n \,=\,2 M_{ab} \,\sin\left(n\,\frac{\pi\,\xi_a}{2}\right) \,\,\,. 
\EN 
As we have previously discussed, this means that, according to the value of $\xi_a$, 
we can have only the following situations at the vacuum $\mid a \,\rangle$: (a) no bound state if $ \xi_a > 1$; (b)  one particle if $\frac{1}{2} < \xi_a < 1$; (c) two particles 
if $\frac{1}{3} < \xi_a < \frac{1}{2}$; (d) $\left[\frac{1}{\xi_a}\right]$ particles 
if $\xi_a < \frac{1}{3}$, although only the first two are stable, the others being 
resonances. So, semiclassically, each vacuum of the theory cannot have more than two stable particles above it. Viceversa, if $\omega = 0$, there are no poles in the Fourier transform of the kink and therefore there are no neutral particles near the vacuum $\mid a \,\rangle$.  
 
\subsection{Asymmetric wells}\label{phi6section}

A polynomial potential with two asymmetric wells must necessarily employ higher powers than $\varphi^4$, and the simplest example is given by a polynomial of maximum power $\varphi^6$. Beside its simplicity, the $\varphi^6$ theory is relevant for the class of universality of the Tricritical Ising Model, and the information available on this model turn out to be a nice confirmation of the semiclassical scenario discussed below. 

\begin{figure}[t]
\centerline{\scalebox{0.4}
{\includegraphics{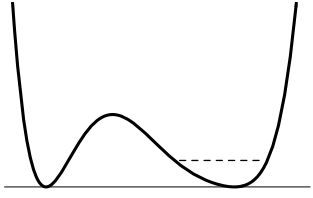}}} 
\caption{{\em Example of $\varphi^6$ potential with two asymmetric wells and a bound state only on one of them.}}
\label{potential6}
\end{figure}

\begin{figure}[b]
\centerline{\scalebox{0.5}
{\includegraphics{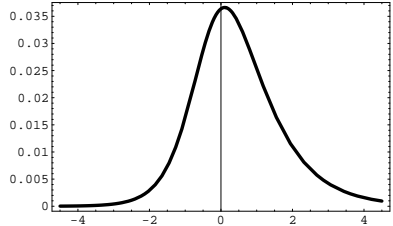}}} 
\caption{{\em Typical shape of $\left(\frac{d \varphi}{dx}\right)_{01}$, obtained by a numerical solution of eq.\,
(\ref{kinkphi6}).}}
\label{asymmsol}
\end{figure}

A class of potentials with two asymmetric wells is given by 
\EQ
U(\varphi) \,=\,\frac{\lambda}{2}\,\left(\varphi + a\frac{m}{\sqrt{\lambda}}\right)^2 \,
\left(\varphi - b\frac{m}{\sqrt{\lambda}}\right)^2 \,\left(\varphi^2 + c \frac{m^2}{\lambda}\right) \,\,\,, 
\label{phi6}
\EN 
with $a,b,c$ all positive numbers. To simplify the notation, it is convenient to use the 
dimensionless quantities obtained by rescaling the coordinate as $x^{\mu} \rightarrow  m x^{\mu}$ and the field as $\varphi(x) \rightarrow \sqrt{\lambda}/m \varphi(x)$, so that  the lagrangian of the model becomes 
\EQ
{\cal L} \,=\,\frac{m^6}{\lambda^2} \left[\frac{1}{2} (\partial \varphi)^2 - 
\frac{1}{2} (\varphi+a)^2 (\varphi-b)^2 
(\varphi^2 + c) \right]\,\,\,.
\label{newphi6}
\EN 
The minima of this potential are localised at $\varphi_0^{(0)} = - a$ and $\varphi_1^{(0)} = b$ and the corresponding ground states will be denoted by $\mid 0 \,\rangle$ and $\mid 1 \,\rangle$, with the curvature of the potential at these points given by 
\EQ
\begin{array}{lll}
U''(-a) & \equiv & \omega^2_0 = (a+b)^2 (a^2 + c) \,\,\,;\\
U''(b) & \equiv & \omega^2_1 = (a+b)^2 (b^2 + c)\,\,\,.
\end{array}
\label{curvature}
\EN 
For $ a \neq b$, we have two asymmetric wells, as shown in Figure \ref{potential6}. Let's assume that the curvature at the vacuum $\mid 0\,\rangle$ is higher than the 
one at the vacuum $\mid 1\,\rangle$, i.e. $a > b$. 
The problem we would like to examine is whether the spectrum of the neutral particles $\mid B \,\rangle_{s}$ ($ s = 0,1$) may be different at the two vacua, in particular, whether it would be possible that one of them (say $\mid 0 \rangle$) has no neutral excitations, whereas the other has just one neutral particle. The ordinary perturbation theory shows that both vacua has neutral excitations, although 
with different value of their mass: 
\EQ
m^{(0)} \,= \,(a+b) \sqrt{2 \, (a^2 + c)} 
\,\,\,\,\,\,\,
,
\,\,\,\,\,\,\,
m^{(1)} \,=\, (a+b) \sqrt{2 \, (b^2 + c)} 
\,\,\,.
\label{baremasses}
\EN 
Let's see, instead, what is the semiclassical scenario. The kink equation is given in this case by 
\EQ
\frac{d\varphi}{d x} \,=\,\pm (\varphi + a) (\varphi - b) \,\sqrt{\varphi^2 + c}\,\,\,.
\label{kinkphi6}
\EN 
Even in the absence of an exact solution, we can extract the main features of this equation. The kink solution interpolates between the values $-a$ (at $ x = -\infty$) and $b$ (at $x = +\infty$), while anti-kink solution does viceversa, but with an important difference: its behaviour at $x = -\infty$ is different from the one of the kink. As a matter of fact, the behaviour at $x = - \infty$ of the kink is always equal to the behaviour at $x = +\infty$ of the anti-kink (and viceversa), but the two vacua are approached, in this theory, differently. This is explicitly shown in Figure \ref{asymmsol} and proved  
in the following. 

Let us consider the limit $x \rightarrow - \infty$ of the kink solution. For these large values of $x$, we can approximate eq.\,(\ref{kinkphi6}) by substituting, in the second and in the third term of the right-hand side, $\varphi \simeq -a$, with the result 
\EQ
\left(\frac{d\varphi}{dx}\right)_{0,1} \simeq (\varphi + a) (a + b) \sqrt{a^2 + c} 
\,\,\,\,\,
,
\,\,\,\,\,
x \rightarrow - \infty
\EN 
This gives rise to the following exponential approach to the vacuum $\mid 0 \rangle$ 
\EQ
\varphi_{0,1}(x) \simeq -a + A \exp(\omega_0 x) 
\,\,\,\,\,\,\,
,
\,\,\,\,\,\,\,
x \rightarrow - \infty 
\EN 
where $A > 0$ is a arbitrary costant (its actual value can be fixed by properly solving the non-linear differential equation). To extract the behavior at $x \rightarrow -\infty$ of the anti-kink, we substitute this time $\varphi \simeq b$ into the first and third term of the right hand side of (\ref{kinkphi6}), so that 
\EQ
\left(\frac{d\varphi}{d x}\right)_{1,0} \simeq (\varphi - b) (a + b) \sqrt{b^2 + c} 
\,\,\,\,\,\,
,
\,\,\,\,\,\,
x \rightarrow - \infty
\EN 
This ends up in the following exponential approach to the vacuum $\mid 1 \rangle$ 
\EQ
\varphi_{1,0}(x) \simeq b - B \exp(\omega_1 x) 
\,\,\,\,\,\,\,\,
,
\,\,\,\,\,\,\,\,
x \rightarrow - \infty
\EN 
where $B > 0$ is another constant. Since $\omega_0 \neq \omega_1$, the asymptotic behaviour of the two solutions gives rise to the following poles in their Fourier transform 
\begin{eqnarray}
{\cal F}(\varphi_{0,1}) & \rightarrow & \frac{A}{\omega_0 + i k} \nonumber \\
& & \label{polephi6}\\
{\cal F}(\varphi_{1,0}) & \rightarrow & \frac{-B}{\omega_1 + i k} \nonumber 
\end{eqnarray}
In order to locate the pole in $\theta$, we shall reintroduce the correct units. Assuming to have solved the differential equation (\ref{kinkphi6}), the integral of its energy density gives the common mass of the kink and the anti-kink. In terms of the constants in front of the Lagrangian (\ref{newphi6}), its value is given by 
\EQ
M\,=\,\frac{m^5}{\lambda^2} \,\alpha \,\,\,,
\EN 
where $\alpha$ is a number (typically of order $1$), coming from the integral of the adimensional energy density (\ref{integralmass}). Hence, the first pole\footnote{In order to determine the others, one should look for the subleading exponential terms of the solutions.} of the Fourier transform of the kink and the antikink solution are localised at 
\begin{eqnarray}
\theta^{(0)} \,& \simeq & \,i\pi \left( 1 - \omega_0 \,\frac{m}{\pi M}\right) = i\pi 
\left(1 - \omega_0 \,\frac{\lambda^2}{\alpha m^4}\right)
\nonumber \\
& & \\
\theta^{(1)} \,& \simeq & \,i\pi \left( 1- \omega_1 \,\frac{m}{\pi M}\right) 
\,=\, i\pi \left(1 - \omega_1 \,\frac{\lambda^2}{\alpha m^4}\right) \nonumber 
\end{eqnarray}
If we now choose the coupling constant in the range 
\EQ
\frac{1}{\omega_0} < \frac{\lambda^2}{m^4} < \frac{1}{\omega_1}    \,\,\,,
\label{range}
\EN 
the first pole will be out of the physical sheet whereas the second will still remain inside it!  Hence, the theory will have only one neutral bound state, localised at the vacuum $\mid 1 \,\rangle$. This result may be expressed by saying that the appearance of a bound state depends on the order in which the topological excitations are arranged: an antikink-kink configuration gives rise to a bound state whereas a kink-antikink does not. 

Finally, notice that the value of the adimensional coupling constant can be chosen so that the mass 
of the bound state around the vacuum $\mid 1 \,\rangle$ becomes equal to mass of the kink. 
This happens when 
\EQ
\frac{\lambda^2}{m^4} \,=\,\frac{\alpha}{3 \omega_1} \,\,\,.
\EN  

Strange as it may appear, the semiclassical scenario is well confirmed by an explicit example. This is provided by the exact scattering theory of the Tricritical Ising Model perturbed by its sub-leading magnetization. Firstly discovered through a numerical analysis of the spectrum of this model \cite{LMC}, its exact scattering theory has been discussed later in \cite{CKM}. 

\section{Conclusions}

In this paper we have shown that quantum integrable models have enough set of constraints to allow the determination of their mass spectrum and correlation functions. Quantum Ising model and Lieb-Liniger model have been our paradigmatic examples to illustrate several aspects of quantum integrability. We have discussed how to control the breaking of integrability by means of two approaches: the Form Factor Perturbation Theory and the semiclassical method. When the breaking of integrability is realized by a non-local operator, one gets the confinement phenomena of the topological excitations.   
Form Factor Perturbation Theory permits also to control the decay processes and computation of the life-time of unstable particles. In addition, we have used simple arguments of the semi-classical analysis to investigate the spectrum of neutral particles in quantum field theories with kink excitations. We have analyzed, in particular, 
two cases: the first relative to a potential with symmetric wells, the second concerning with a potential with asymmetric wells. As a general result of this analysis, we have the existence of a critical value of the coupling constant, beyond which there are no neutral bound states. Another result is about the maximum number $n \leq 2$ of neutral particles above a generica vacuum of a non-integrable theory. An additional aspect is the role played by the asymmetric vacua, which may have a different number of neutral excitations above them. Let's finally mention that integrable and non-integrable models can be numerically studied by means of the so-called Truncated Conformal Space Approach \cite{alyoshatruncation}, underlying for instance the different statistics which rules the energy level difference, as shown in \cite{Brandino}

\end{document}